\documentclass[useAMS,usenatbib]{mnras}
\usepackage{amsmath}
\usepackage{amssymb}
\usepackage{float}
\usepackage{graphicx}

\newcommand{\vect}[1]{\mathbfit{#1}}

\newcommand{\unit}[1]{\, \mathrm{#1}}
\newcommand{\subscript}[1]{_{\mathrm{#1}}}

\newcommand{\fn}[2]{\, \mathrm{#1} \,{#2}}

\newcommand{\sectionref}[1]{\hyperref[#1]{Section~\ref*{#1}}}
\newcommand{\autorefp}[1]{(\autoref{#1})}

\newif\ifdraft

\drafttrue
\draftfalse

\ifdraft
\newcommand{\columnfigure}[3]{\begin{figure}\caption{#2}\label{#3}\end{figure}}

\newcommand{\triplefigure}[5]{\begin{figure*}\caption{#4}\label{#5}\end{figure*}}
\newcommand{\linefigure}[3]{\begin{figure*}\caption{#2}\label{#3}\end{figure*}}

\newcommand{\scalelinefigure}[4]{\begin{figure*}\caption{#3}\label{#4}\end{figure*}}
\newcommand{\towerfigure}[4]{\begin{figure}\caption{#3}\label{#4}\end{figure}}
\else
\newcommand{\columnfigure}[3]{\begin{figure}\includegraphics[width=\columnwidth]{#1}\caption{#2}\label{#3}\end{figure}}

\newcommand{\triplefigure}[5]{\begin{figure*}\includegraphics[width=0.33\linewidth]{#1}\includegraphics[width=0.33\linewidth]{#2}\includegraphics[width=0.33\linewidth]{#3}\caption{#4}\label{#5}\end{figure*}}
\newcommand{\linefigure}[3]{\begin{figure*}\includegraphics[width=\linewidth]{#1}\caption{#2}\label{#3}\end{figure*}}

\newcommand{\scalelinefigure}[4]{\begin{figure*}\begin{center}\includegraphics[width=#1\linewidth]{#2}\caption{#3}\label{#4}\end{center}\end{figure*}}
\newcommand{\towerfigure}[4]{\begin{figure}\begin{flushright}\includegraphics[width=0.9\columnwidth]{#1} \\ \includegraphics[width=\columnwidth]{#2}\end{flushright}\caption{#3}\label{#4}\end{figure}}
\fi

\title[Small-scale dynamo in galaxies - saturation phase]{A small-scale dynamo in feedback-dominated galaxies - II. The saturation phase and the final magnetic configuration}
\author[M. Rieder, R. Teyssier]{Michael Rieder\thanks{Contact e-mail: \href{rieder@physik.uzh.ch}{rieder@physik.uzh.ch}} and Romain Teyssier\\
Institute for Computational Science\\
Centre for Theoretical Astrophysics and Cosmology\\ 
Universit\"at Z\"urich, 8057 Z\"urich, Switzerland}

\begin{document}

\ifdraft
\else
\maketitle
\fi

\begin{abstract}
Magnetic fields in galaxies are believed to be the result of dynamo amplification of initially weak seed fields, reaching equipartition strength inside the interstellar medium. 
The small-scale dynamo appears to be a viable mechanism to explain observations of strong magnetic fields in present-day and high-redshift galaxies, 
considering the extreme weakness of seed fields predicted by battery mechanisms or primordial fields. 

Performing high-resolution adaptive mesh magneto-hydrodynamic simulations of a small mass, isolated cooling halo with an initial magnetic seed field strength well below equipartition, 
we follow the small-scale dynamo amplification from supernova-induced turbulence up to saturation of the field. 
We find that saturation occurs when the average magnetic pressure reaches only 3 \% to 5 \% of the turbulent pressure. 
The magnetic energy growth transitions from exponential to linear, and finally comes to halt. 
The saturation level increases slightly with grid resolution. 
These results are in good agreement with theoretical predictions for magnetic Prandtl numbers of order $\mathrm{Pr_M} \sim 1$ and turbulent Mach numbers of order $\mathrm{M} \sim 10$. 
When we suppress supernova feedback after our simulation has reached saturation, 
we find that turbulence decays and that the gas falls back onto a thin disk with the magnetic field in local equipartition. 

We propose a scenario in which galactic magnetic fields are amplified from weak seed fields in the early stages of the Universe to sub-equipartition fields, 
owing to the turbulent environment of feedback-dominated galaxies at high redshift, and are evolved further in a later stage up to equipartition, 
as galaxies transformed into more quiescent, large spiral disks.
\end{abstract}


\section{Introduction}

Measurements of Faraday rotation in the Milky Way \citep{taylor2009rotation}, in nearby galaxies \citep{2016A&ARv..24....4B} as well as in high-redshift galaxies \citep{M-2008Natur.454..302B} reveal strong magnetic fields, usually close to equipartition with the turbulent energy density. 
\cite{2008ApJ...680..981R} have detected field strengths up to $18 \unit{mG}$ in starburst galaxies  but ordered galactic magnetic fields in the ISM of normal spiral galaxies are typically of the order of several $\unit{\mu G}$.
Their field lines mostly exhibit a spiral structure if the galaxy is itself a grand design spiral galaxy \citep{2013pss5.book..641B} but, interestingly, this can also be the case for ring galaxies like NGC 4736 \citep{2008ApJ...677L..17C}, flocculent galaxies like NGC 4414 without clear spiral arms \citep{2002A&A...394...47S} or in the central regions of galaxies. 
Strong ordered fields are found at the edges of optical arms with dense cold molecular gas in M 51 \citep{2006A&A...458..441P} but can also form their own magnetic arms not coinciding with the gaseous or the optical spiral arms like in NGC 6946 \citep{2007A&A...470..539B}.
In a set of aggregated data on 20 spiral galaxies from the literature, \cite{2015ApJ...799...35V} report pitch angles ranging between $-8\unit{^\circ}$ and $-48\unit{^\circ}$ with a mean value of $-25\unit{^\circ}$ and find a correlation between the spiral arm pitch angles and the magnetic pitch angles.

The origin of magnetic fields in the Universe might be primordial \citep{2013A&ARv..21...62D} or due to microphysical processes at later epochs, such as the Biermann battery \citep{M-1950ZNatA...5...65B} in shock fronts \citep{0004-637X-480-2-481} or ionization fronts \citep{0004-637X-539-2-505}, 
spontaneous fluctuations \citep{2012PhRvL.109z1101S} or fluctuations due to the Weibel instability \citep{2009ApJ...693.1133L} in the plasma of protogalaxies,
or even magnetic fields released into the ISM by stars through stellar winds or supernova outbursts \citep{1973SvA....17..137B} 
or even by AGN jets \citep{2005LNP...664....1R} and subsequently diluted. 

Microphysical mechanisms (such as the Biermann battery) are capable of creating magnetic fields of the order of $10^{-20} \unit{G}$, 
while the constraints on the primordial field are less definite because the difference between lower and upper limits remains vast. 
Upper limits can be derived from Big Bang Nucleosynthesis (BBN) abundances and from the large-scale density structure or the Cosmic Microwave Background (CMB). 
\cite{2016A&A...594A..19P} currently set the upper limit of the primordial magnetic field (PMF) field strength to $10^{-9} \unit{G}$ in the comoving frame based on their CMB anisotropy measurements. 
$\gamma$-ray observations of blazar spectra give lower limits for the field strength ranging from $10^{-18} \unit{G}$ up to $10^{-15} \unit{G}$ \citep{2010Sci...328...73N,2011ApJ...733L..21D,2012ApJ...747L..14V}, based on the remnants of the PMF that are believed to thread the intergalactic medium. This leaves us to explain many orders of magnitude magnetic field amplification in a timeframe of just a few Gyr.

Large-scale dynamos (LSD) are a viable mechanism to amplify magnetic fields coherently on (as the name suggests) large spatial scales. 
Theoretical models for galactic LSD exhibit exponential growth rates of the order of $\Gamma \simeq 0.01 - 0.1 \Omega$ \citep{pariev2007magnetic}, 
where $\Omega$ is the galactic angular rotation rate. This translates into an e-folding time scale of roughly 1~Gyr in a typical present-day spiral, 
making it virtually impossible to amplify the field as fast as required by the observations. 
\cite{1992ApJ...401..137P} proposed also a cosmic-ray-driven LSD, that was simulated for the first time by \cite{2004ApJ...605L..33H}, 
who found a larger growth rate of $\Gamma \simeq \Omega$.

Small-scale dynamos (SSD) on the other hand, can have very fast magnetic field amplification, 
with timescales of the order of the eddy turnover times of the smallest turbulent eddies \citep{2012SSRv..169..123B}. 
The theoretical foundation of this dynamo theory is commonly attributed to \cite{kazantsev1968}. 
\cite{M-1992ApJ...396..606K} considered the Kolmogorov power spectrum of small-scale velocity fluctuations for the galactic dynamo 
and found that the magnetic energy spectrum scales with the wavenumber as $k^{3/2}$ on scales larger than the resistive scale. 
\cite{2015PhRvE..92b3010S} presented a theoretical analysis of the SSD, in the limits of both small and large magnetic Prandtl numbers, 
finding that its growth rate scales with respectively the magnetic or kinetic Reynolds number. 
They also evaluated the ratio of magnetic to turbulent energy after saturation, finding values between $0.1 \%$ and $50\%$, 
depending on the model of turbulence, on the value of Pr$_M$ and on the value of the Mach number. 
Their results are confirmed by previous numerical investigations on the turbulence-driven dynamo such as \cite{2011PhRvL.107k4504F,2014ApJ...797L..19F} or \cite{2016MNRAS.461.1260T},
who also reported  that the saturation level is slightly increasing with resolution. 
The main issue with the SSD mechanism is, however, that it creates strong fluctuating fields, for which the large scale component is subdominant, 
and generally does not reach equipartition. These two properties are in contradiction with observational data of nearby galaxies \citep{2015A&A...578A..93B}. 
On the other hand, it is plausible that the magnetic fields we observe in galaxies are not the result of just one single process, 
but probably a combination of various mechanisms, such as the two dynamos theory \citep{2015PhRvL.115q5003S} 
or various reconnection processes during the hierarchical assembly of galaxies.

The importance of feedback processes has been recognised increasingly over the past decade in galaxy formation theory \citep{2012MNRAS.tmp.2970S}, 
along with the influence of associated galactic winds \citep{2006MNRAS.373.1265O} 
and the dominant cold stream accretion mechanism \citep{2005MNRAS.363....2K,2008MNRAS.390.1326O,2009Natur.457..451D}. 
Observations suggest that galactic winds are ubiquitous in star bursting local galaxies \citep{1999ApJ...513..156M},
as well as many ``normal'' high redshift galaxies \citep{2010ApJ...717..289S}. 
Abundance matching between dark matter haloes and observed central galaxies \citep{M-2013ApJ...770...57B,M-2013MNRAS.428.3121M} suggests considerably stronger feedback processes than previously considered for early galaxies to regulate star formation through cosmic evolution, 
especially at high redshift, to maintain such a low  star formation efficiency \citep{2013ApJ...770...25A,Hopkins:2014ua,2014MNRAS.444.2837R,Wang:2015dc}. 
Interestingly, \cite{2016ApJ...829..133K} found a correlation between strong magnetic field signatures in the Faraday Depth spectrum and strong Mg II absorption, 
which is associated with superwinds from starbursting galaxies \citep{0004-637X-562-2-641}, suggesting a link between strong outflows and a high magnetisation of the intergalactic medium.

In this rather violent, feedback-dominated scenario, dwarf galaxies play a very important role. 
They are the dominant galaxy population at high redshift, probably responsible for the cosmic re-ionisation \citep{2014ApJ...788..121K}. 
They are also the progenitors of the Milky Way satellites, which are useful laboratories to test our current galaxy formation paradigm. 
For the latter, violent feedback mechanisms have also been invoked to explain the absence of cusp in the dark matter density profile, 
and the presence of a dark matter core in low surface brightness galaxies \citep{2001ApJ...552L..23D}. 
Cosmological simulations of dwarf galaxies have been performed with strong feedback recipes, 
confirming in this case the formation of a dark matter core \citep{2010Natur.463..203G,2012MNRAS.422.1231G,2013MNRAS.429.3068T}. 

In galaxies, the velocity field on both small and large scales resulting from repeated giant feedback events can have a direct influence on the growth of the magnetic energy. 
Indeed, supernovae explosions in the Milky Way have been considered for quite a long time as a source of helical gas motions, 
promoting a large-scale $\alpha$-dynamo in the Galaxy \citep{1992ApJ...391..188F}. 
The Milky Way is a rather quiescent galaxy with moderate supernova activity but might have undergone different phases in its lifetime. 
We therefore want to study in this work the small-scale dynamo-induced growth of magnetic fields in dwarf galaxies under the influence of strong feedback, 
how it reaches saturation and how the magnetic field could evolve when feedback becomes weaker. 
In the first step, we are considering a feedback-dominated galaxy, with a high star formation rate and violent turbulent motions, 
together with large-scale galactic fountains or winds. As the magnetic field grows until the small-scale dynamo is saturated, 
this feedback is turned off in the second step to see how it evolves in a less turbulent, quiescent phase.

In recent years, several simulations of the magnetic fields evolution have been performed  in the context of galaxy formation \citep{M-2009ApJ...696...96W,2010A&A...523A..72D}, 
These early studies were based on the  ``cooling halo'' numerical set-up, and have achieved only moderate magnetic field amplification. 
An important shortcoming in these simulations was the absence of feedback \citep{M-2009ApJ...696...96W}, 
or the relative weakness of the feedback recipe used at that time \citep{2010A&A...523A..72D}. 
\cite{2012MNRAS.422.2152B} performed a simulation using a new developed MHD version of the GADGET code GADGET, using a divergence cleaning technique, 
and they observed a fast exponential growth of the magnetic field, which they attributed to a small-scale dynamo. 
Surprisingly, they did not include any explicit source of turbulence resulting in a relatively smooth flow, but reported nevertheless a very large growth rate. 

More recently, \cite{M-2013MNRAS.432..176P} also observed strong magnetic field amplification using the MHD version of the AREPO code \citep{2011MNRAS.418.1392P}, 
although also in this case, stellar feedback effects were not introduced explicitly, and the smoothness of their effectively 2D rotating flow would make dynamo amplification notoriously difficult to obtain. 
The same code was used very recently for cosmological zoom-in simulations of Milky Way-like disk galaxies in \cite{2017arXiv170107028P}, 
this time including a more realistic treatment of stellar feedback, resulting in strong turbulence driving.  
As a consequence, a fast magnetic energy amplification was observed at high redshift, attributed to the SSD, followed by a slower amplification at lower redshift, attributed to the LSD. 

In our previous paper \citep{2016MNRAS.457.1722R}, simulating a dwarf galaxy at high resolution, we found strong evidence for a small-scale dynamo operating in galaxies with feedback-driven turbulence, with e-folding timescales of up to 100~Myr. A similar approach was used in \cite{2016arXiv161008528B} for a Milky Way galaxy where the magnetic field was seeded by supernova ejections. Another interesting study was conducted recently by \cite{2016MNRAS.461.4482D} who impose a spiral potential in their simulations of an isolated disk and found magnetic field reversals.

In the present paper, we follow up on our previous work \citep{2016MNRAS.457.1722R}, using again the isolated cooling halo setup, 
with strong stellar feedback to investigate the mechanism of magnetic field amplification through the SSD. 
We use the Adaptive Mesh Refinement code RAMSES \citep{2002AA...385..337T}, adopting the ``Constrained Transport'', 
strictly divergence-free-preserving, MHD solver presented in \cite{2006JCoPh.218...44T} and in \cite{2006AA...457..371F}.
We focus here our analysis of the saturation properties of the SSD in the galaxy formation context. 
The paper is organised as follows: In \sectionref{chap:num-methods}, we present our numerical methods in terms of galaxy formation physics and magnetic fields evolution, 
as well as the intricacies of modelling realistic turbulent flows in numerical simulations. 
In \sectionref{chap:IC}, we describe our initial conditions for the isolated, magnetised cooling dwarf halo. 
In \sectionref{chap:analysis}, we present our main results on dynamo saturation and study the impact of resolution and stellar feedback. 
In \sectionref{chap:discussion}, we discuss our results in the context of current galactic dynamo theories. 
Finally, in \sectionref{chap:conclusions}, we discuss the implications of our work for our understanding of cosmic magnetism.

\section{Numerical Methods}
\label{chap:num-methods}

We use the Adaptive Mesh Refinement (AMR) code RAMSES \citep{2002AA...385..337T} to simulate the formation and evolution of an isolated dwarf galaxy. 
It is modelled as a magnetised ideal plasma, coupled through gravity to a collisionless fluid made of dark matter and stars, 
with additional numerical schemes to account for physical sub-resolution processes such as gas cooling, star formation and supernova feedback. 
In this section, we outline the numerical methods used in this work to model self-gravitating supersonic turbulence driven by stellar feedback.

\subsection{Ideal MHD with Gravity}
We solve the ideal MHD equations which are written here without gravity and cooling source terms for the sake of simplicity:
\begin{align}
\partial_{t} \rho + \nabla \cdot (\rho \vect u) &= 0 \\
\partial_{t} (\rho \vect u) + \nabla \cdot ( \rho \vect u \vect u^{T} - \vect B \vect B^{T} + P \subscript{tot} ) &= 0 \\
\partial_{t} E + \nabla \cdot \left[ (E + P \subscript{tot}) \vect u - (\vect u \cdot \vect B) \vect B \right] &= 0 \\
\partial_{t} \vect B - \nabla \times \left( \vect u \times \vect B \right) &= 0
\label{induction-eqn}
\end{align}
where $\rho$ is the gas density, $\rho \vect u$ is the momentum, $\vect B$ is the magnetic field, $E = \frac 1 2 \rho \vect u^{2} + \rho \varepsilon + \frac 1 {2} \vect B^{2}$ is the total energy, and $\varepsilon$ is the specific internal energy. The total pressure is given by $P_{tot} = P + 
\frac{1}{2} \vect B^{2}$ where we assume a perfect gas equation of state $P = (\gamma - 1) \rho \varepsilon$. 
This system of conservation laws is furthermore completed by the solenoidal constraint
\begin{equation}\label{solenoidal} \nabla \cdot \vect B =0.\end{equation}

RAMSES uses a hybrid approach with gas variables stored on a tree-based adaptively refined mesh, while dark matter and stars are tracked by collissionless particles. The equations above are solved using the second-order unsplit Godunov scheme based on the MUSCL-Hancock method with the HLLD Riemann solver and MinMod slope limiter. The induction equation (\autoref{induction-eqn}) is solved with the Constrained Transport (CT) method \citep{2006JCoPh.218...44T}, which preserves the divergence of the magnetic field $\nabla \cdot \vect B =0$ from the initial conditions, and MonCen sloper limiter. We use free-outflow boundary conditions with imposed zero-gradient at the simulation box boundaries for the gas variables and the perpendicular magnetic field. 

To treat physical processes that are well below the resolution limit but nevertheless important in the theory of galaxy formation, 
we include several effects such as gas cooling, star formation and supernova feedback. Gas cooling is implemented using a standard H and He cooling function, with an additional metal cooling component, 
as in \cite{1993ApJS...88..253S} for temperatures above $10^{4}$ K, 
and fine-structure cooling of [CI] and [OII] below $10^4$~K, based on \cite{1995ApJ...440..634R}. 
Cooling stops at a minimum pressure floor in order to ensure the Truelove criterion \citep{1997ApJ...489L.179T} and always resolve the Jeans length by at least 4 grid cells. 
We create star particles as a random Poisson process compliant with a Schmidt law as in \cite{2006A&A...445....1R}. 
The effect of supernovae is modelled by releasing non-thermal energy into the ISM over a dissipation time scale of 20~Myr \citep{2013MNRAS.429.3068T}. 
We refer the interested reader to \cite{2016MNRAS.457.1722R} for a more detailed account.

\subsection{Turbulence}

Modelling turbulence properly is central to our goal of simulating the small-scale dynamo. 
Two key quantities in this context are the Reynolds number, which is the ratio of inertial vs. viscous forces
\begin{equation}\mathrm{Re} = \frac{u L}{\nu}\end{equation}
where $\nu$ is the kinematic viscosity and $u$ and $L$ are the typical velocity and length scale of the problem, and its counterpart for the magnetic field, the magnetic Reynolds number as the ratio of induction vs. diffusion
\begin{equation}\mathrm{Re_M} = \frac{u L}{\eta}\end{equation}
where $\eta$ is the magnetic diffusitivity. Additionally, it is useful to define the magnetic Prandtl number
\begin{equation}\mathrm{Pr_M = \frac{Re_M}{Re}} = \frac{\nu}{\eta}\end{equation}
as the ratio of those two quantities and thus as the ratio of kinematic viscous and magnetic diffusivity. 
Our simulations do not account for microscopic diffusion processes, as we aim at solving the ideal MHD equations, 
but the stability of the numerical solution and its convergence towards the weak solution of the underlying model equations 
are both enforced by the numerical diffusion inherent to the Godunov scheme. 
Due to our limited spatial resolution of $\Delta x \simeq 10 \unit{pc}$, 
the numerical effective viscosity and magnetic diffusivity coefficients are both much greater than the typical physical values for the typical ISM. 

RAMSES has been tested to successfully reproduce idealised fast dynamo flows \citep{2006JCoPh.218...44T} such as the ABC flow \citep{1986GApFD..36...53G,1995stf..book.....C} or the Ponomarenko dynamo \citep{1973JAMTP..14..775P}. The numerical magnetic Reynolds number was shown to be proportional to the inverse square of the number of grid points, which is to be expected with second-order schemes and rather smooth solutions. However, in the context of a highly complex flow such our turbulent galaxy, where kinematics is dominated by rotation and supernova explosions inside a deep gravitational potential, it becomes virtually impossible to exactly determine neither viscosity nor diffusion due to the numerical scheme. This is even further complicated by the adaptively refined mesh. Qualitatively, it is sufficient to know that viscous effects as well as magnetic reconnection and diffusion caused by the numerical scheme happen at a length scale which is close to the mesh size, which is mainly the cell size at the maximal refinement level. We also point out that numerical viscosity and numerical diffusion are approximately equal due to their identical numerical origin, so that the effective magnetic Prandtl number $\mathrm{Pr_M} \simeq 1$ \citep[cf.][for more discussion]{2015ARA&A..53..325T}.

\section{Initial Conditions}
\label{chap:IC}

As in the first paper of this series \citep{2016MNRAS.457.1722R}, we simulate an isolated dark matter halo with gas cooling and a small initial rotation. 
In our previous work, we have conducted a study on varying initial conditions with different halo sizes and initial magnetic field topologies 
and we found no significant impact on the nature of the dynamo itself. 
If the feedback mechanism is strong enough to stir strong turbulence in the disk and launch a large scale galactic wind, the magnetic field lines quickly become mangled, 
so that their initial topology and symmetry are lost and the flow develop a strong characteristic random, quasi isotropic and mostly small-scale component. 
We are now especially interested in the final phase of the dynamo mechanism, when the magnetic energy growth reaches saturation.

\begin{table}
\centering
\caption{Parameters of initial halo set-up and additional physics mechanisms}
\label{haloic}
\begin{tabular}{ r | r | l }
\hline
parameter & value & unit \\
\hline
$R_{200}$ & 50 & kpc \\
$V_{200}$ & 35 & km/s \\
$M_{200}$ & 14 & $10^{9}$M$_{\odot}$ \\
$c$ & 10 & \\
$\lambda$ & 0.04 & \\
$f_{\rm gas}$ & 15 & \% \\
$T_{*}$ & 100 & K \\
$\epsilon_{*}$ & 1 & \% \\
$\eta_{\rm SN}$ & 10 & \% \\
$Z_{\rm ini}$ & 0.05 & Z$_{\odot}$\\
metal yield & 10 & \% \\
$B_{\rm max}$ & 0.35 & nG \\
sim. box length & 300 & kpc \\
\hline
\end{tabular}
\end{table}

The initial dark matter and baryonic matter densities follow the NFW profile \citep{1997ApJ...490..493N} with concentration parameter $c=10$ and spin parameter $\lambda = 0.04$ as in \cite{2013MNRAS.429.3068T}. Dark matter is sampled by $10^6$ particles and set to be in equilibrium with the gas by the density-potential pair approach of \cite{2004ApJ...601...37K} and \cite{2006MNRAS.367..387R}. The temperature profile is also initialised to be in hydrostatic equilibrium. The numerical values of the parameters specific to this set-up are given in \autoref{haloic}. In order to study the effect of numerical resolution on the dynamo saturation properties, we increase the resolution in a second simulation run. The resolution parameters for the two runs are given in \autoref{ic-res}.

\begin{table}
\centering
\caption{Size of smallest grid cells and mass resolutions}
\label{ic-res}
\begin{tabular}{ r | c | c | c | c | c }
\hline
resolution & $l_{\rm max}$ & $\Delta$x & $m_{\rm res}$ & $m_{*}$ & $n_{*}$ \\
 &  & [pc] & [M$_{\odot}$] & [M$_{\odot}$] & [H/cc] \\
\hline
low & 14 & 18 & 1523 & 2108 & 14 \\
high & 15 & 9 & 190 & 264 & 112 \\
\hline
\end{tabular}
\end{table}


In \cite{2016MNRAS.457.1722R}, we have studied the dynamo only in the kinematic phase, choosing an arbitrarily small value for the initial magnetic energy.  
Now we want to see what happens when the magnetic field becomes strong enough to become dynamically important. 
We have therefore to increase the initial magnetic field strength, but keeping it small enough for the dynamo to operate in the kinematic phase for some time 
(so that the initial magnetic energy is several orders of magnitude smaller than equipartition), but not too small for the dynamo to saturate in a reasonable amount of time.

Initialising the magnetic field on an adaptively refined grid is a non-trivial task. 
In order to satisfy the zero-divergence constraint \autorefp{solenoidal}, it would be tempting to simply set the initial field to a constant, as
\begin{equation}\vect B\subscript{0} = \begin{pmatrix}0\\0\\B_{0}\end{pmatrix}\end{equation}
throughout the whole simulation box \citep[see e.g.][]{M-2009ApJ...696...96W,M-2013MNRAS.432..176P}. 
However, this approach can lead to some numerical problems, as the Alv\'en wave speed
\begin{equation}v_A = \frac{B}{\sqrt{4\pi\rho}}\end{equation}
becomes very large at the simulation box boundary, where the density falls off by many orders of magnitude, and the magnetic energy accreted at late time will be much larger, 
comparatively to the gas internal or gravitational energy, than the magnetic energy accreted at early time. A constant magnetic field would also not be realistic, 
since a frozen-in magnetic field following the collapse of the hot gas in hydrostatic equilibrium into a dark matter halo should scale as $\left\vert\vect B \right\vert \propto \rho^{2/3}$. 

Combined with the aforementioned solenoidality constraint, this requires a more complex field topology. For this purpose, as in  \cite{2016MNRAS.457.1722R}, 
we define the vector potential
\begin{equation}\vect A\subscript{0} = B_{0}\left[\frac{\rho(r,z)}{\rho_{0}}\right]^{2/3}  r \vect e \subscript{\phi}\end{equation}
where $\rho(r,z)$ is the initial gas density given by the NFW profile and $\vect e \subscript{\phi}$ is the unit vector along the toroidal direction. The initial magnetic field is then set as the curl
\begin{equation}\vect B_0 = \nabla \times \vect A_0,\end{equation}
so that it has exactly zero divergence by design.
The corresponding magnetic field has a vertical component which is symmetric with respect to the mid plane, while its radial component is antisymmetric so that its shape resembles that of a dipole.


\section{Results}
\label{chap:analysis}

\triplefigure{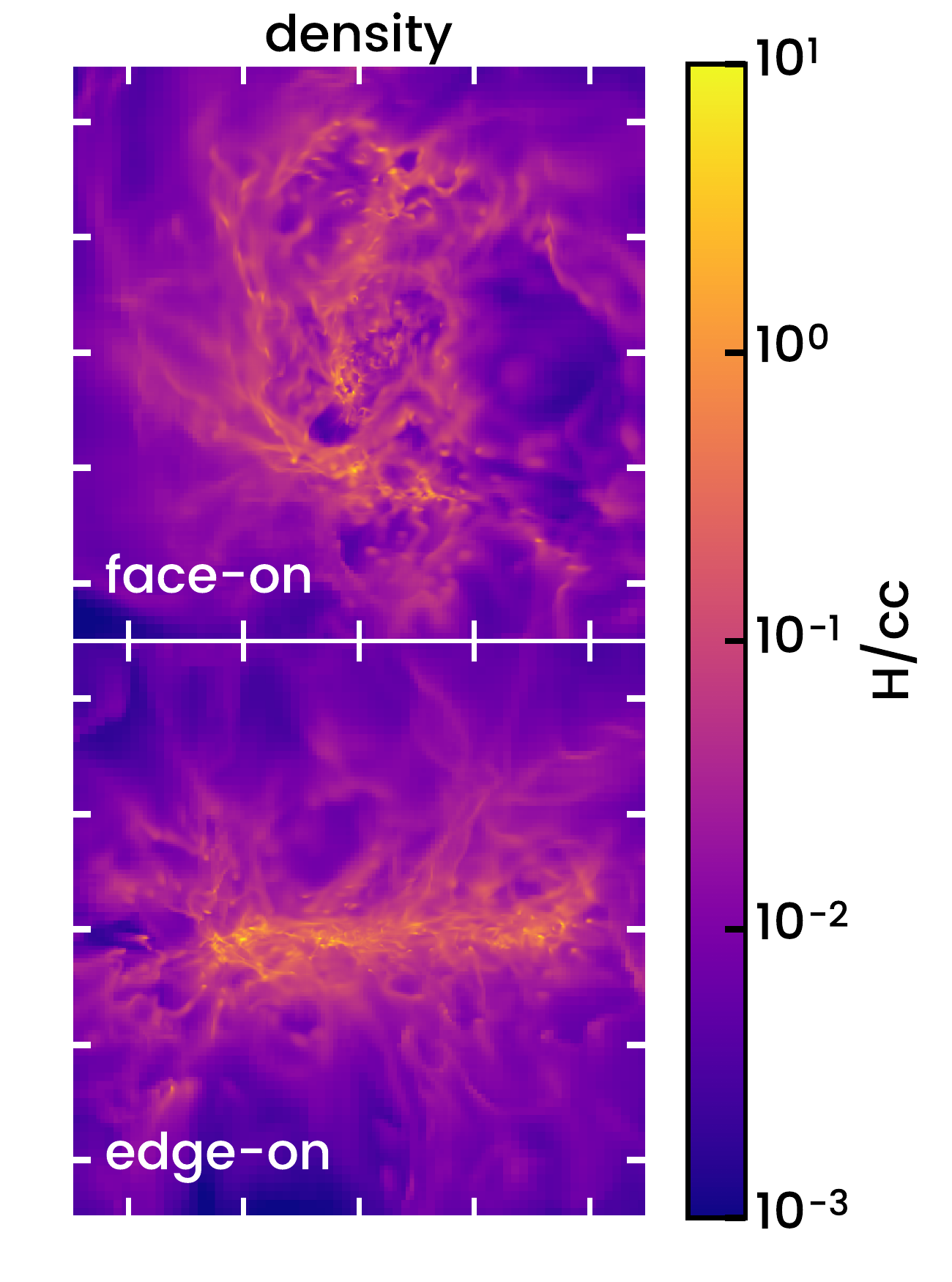}{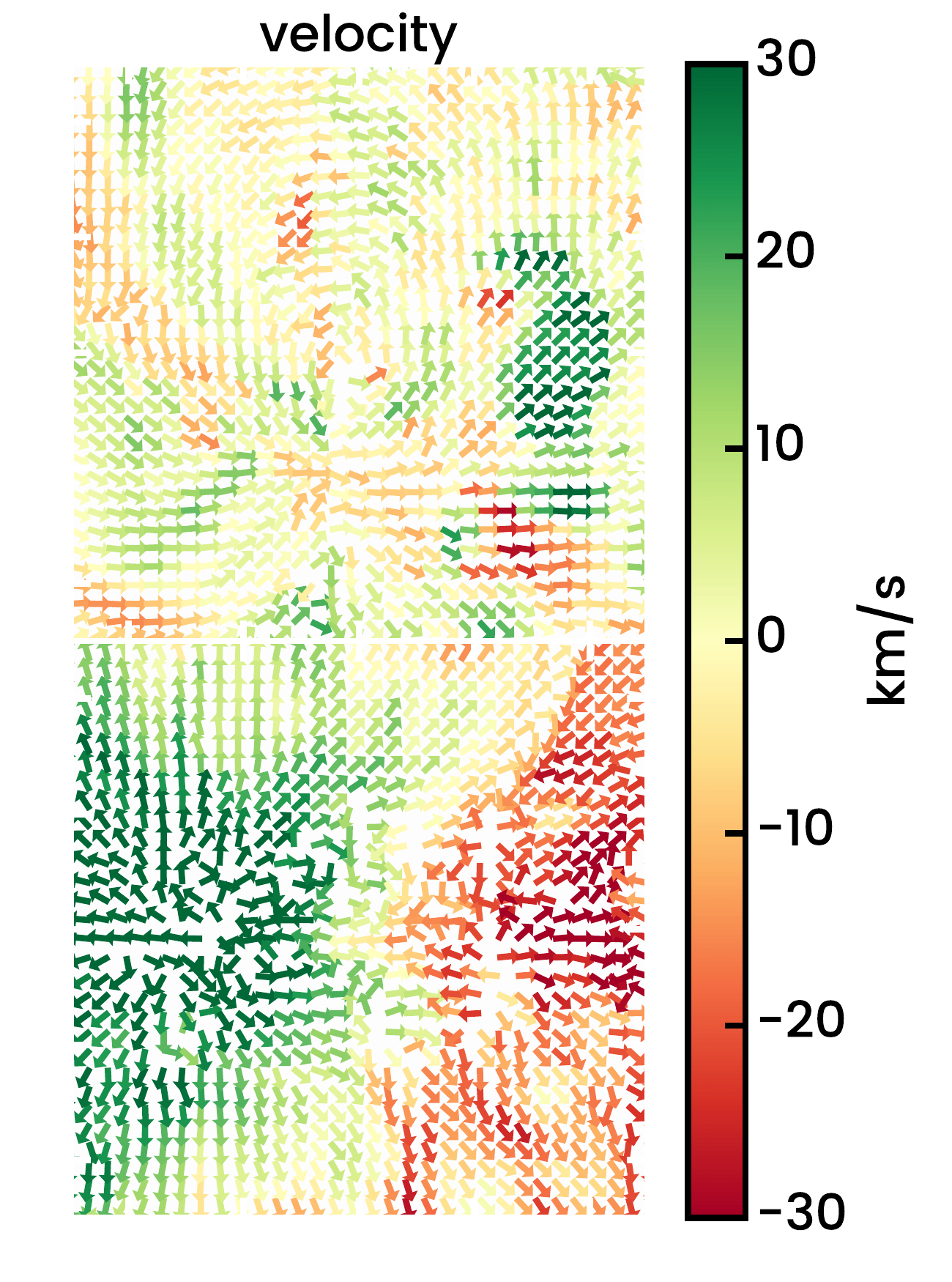}{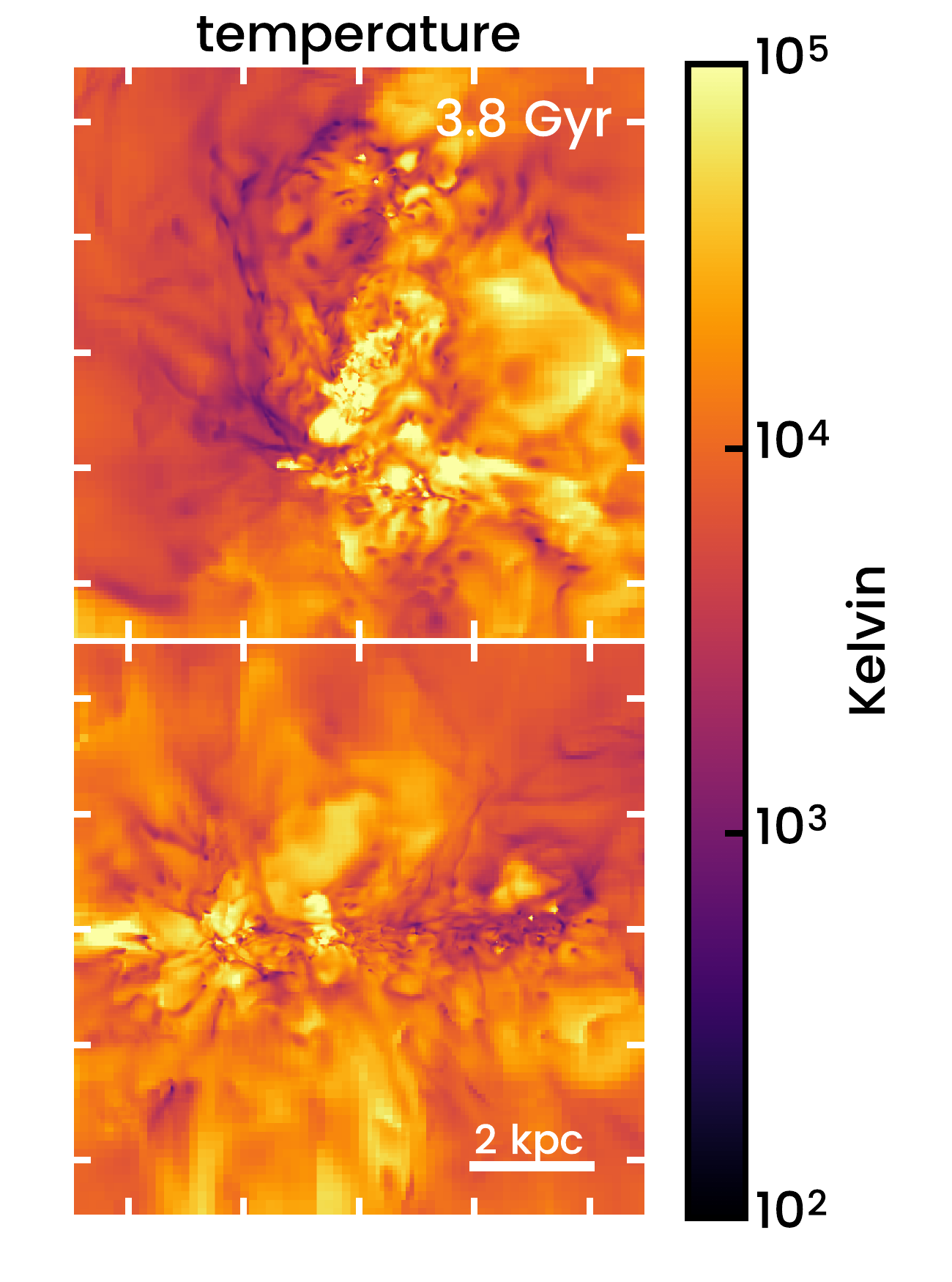}{Line-of-sight projections for different plasma properties looking face-on \textit{(top row)} and edge-on \textit{(bottom row)} in the central 10 kpc cube in the $\Delta x = 18 \unit{pc}$ (lower resolution) simulation at the simulation time $t = 3.8 \unit{Gyr}$. \textit{Left:} density $\langle\rho\rangle$ \textit{Center:} velocity \textit{Right:} temperature}{fig:maps_fbk}

The gas, though initialised in hydrostatic equilibrium, immediately starts to cool radiatively and loses thermal energy. 
The spherical structure collapses onto a rotationally supported disk in only a few Myr. 
This causes the gas disk to fragment into clumps denser than the star formation threshold and the first stars form. 
Their supernova explosions drive outflows that reach out several kpc above and below the galactic midplane 
and quickly, a self-regulated galactic fountain is established, where dissipative processes such as cooling and shocks balance off kinetic energy injection through supernovae explosions. 
The resulting kinematic properties are in very good agreement with those of observed nearby isolated dwarf galaxies \citep{2013MNRAS.429.3068T}. 

This galactic fountains turns the whole galaxy into a giant ``washing machine'' that maintain a very high level of turbulence, injected on very large scales, namely the scale of the entire galaxy. 
Line-of-sight projections of density, velocity and temperature are plotted in \autoref{fig:maps_fbk}. 
The small-scale structure with clumps and filaments is clearly visible. Hot bubbles resulting from SN explosions, with temperature between $10^5$ and $10^6$~K, 
rise until the gas can cool down and falls back to the midplane, so that instead of ordered rotation, we see a highly turbulent velocity field configuration. 

Since the gas is not isothermal, the flow Mach number ${\cal M} = v/c_{\rm s}$, where $v$ is the local gas velocity and $c_s$ is the local sound speed, 
is a non-trivial parameter to determine. 
Nevertheless, we would like to make a good estimate of the typical Mach number in our flow. 
We plot in \autoref{fig:mach} a mass-weighted histogram of the Mach number for every cell.
We find that the flow is highly supersonic, with most of the gas mass at the Mach number in the range of ${\cal M} \approx 6 - 7$.

\columnfigure{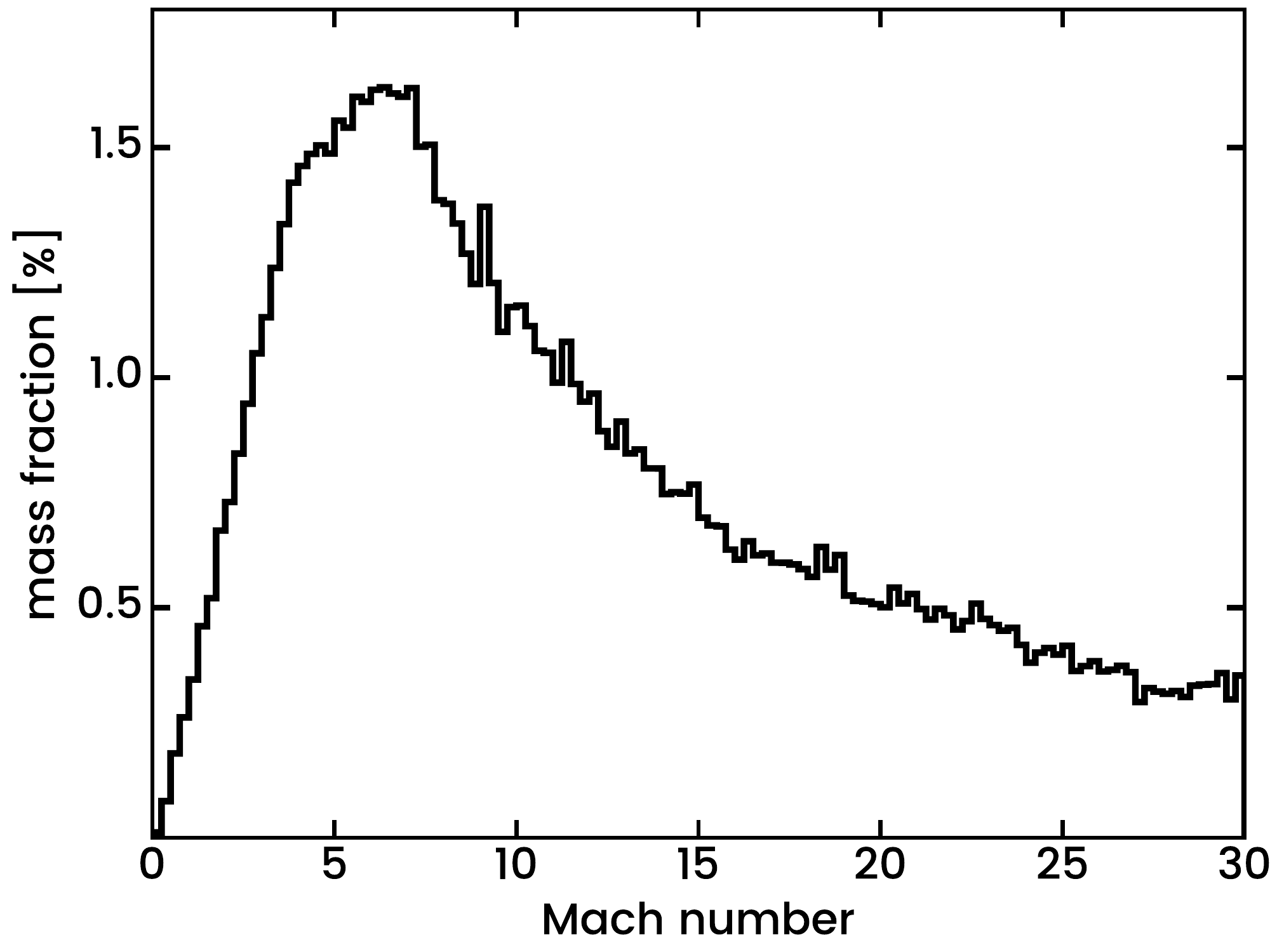}{Mass histogram of local cell Mach number. The flow is highly supersonic with most of the mass at $M \approx 6-7$.}{fig:mach}

\subsection{Saturation of the small-scale dynamo}

As demonstrated in our previous work \citep{2016MNRAS.457.1722R}, the strong turbulence injected on large scale by the galactic fountain  
triggers a small-scale dynamo that amplifies the magnetic field exponentially fast. 
The time evolution of the total magnetic energy inside the whole simulation box is given in \autoref{fig:emag}, using both a linear and a logarithmic scale. 
The initial simulation phase is characterised by an exponential growth, typical of a fast dynamo in the kinematic regime, for which the field strength is too weak to have an effect on the flow.
We measured a best-fit exponential growth rate of $\Gamma \simeq 3.3 \unit{Gyr}^{-1}$. 
This exponential amplification continues until the magnetic field becomes strong enough for the Lorentz force to back-react on the velocity field: the dynamo enters its non-linear phase. 
At that point, which is after $t\subscript{nl} \simeq 2.3 \unit{Gyr}$, the growth rate of magnetic energy becomes weaker and transitions into the onset of the saturation phase, where the growth
is now linear in time, and not exponential anymore. 
Eventually, the field strength becomes so high that the dynamo saturates completely and the field amplification stops, in our simulation after a time of $t\subscript{sat} \simeq 3.8 \unit{Gyr}$.

\columnfigure{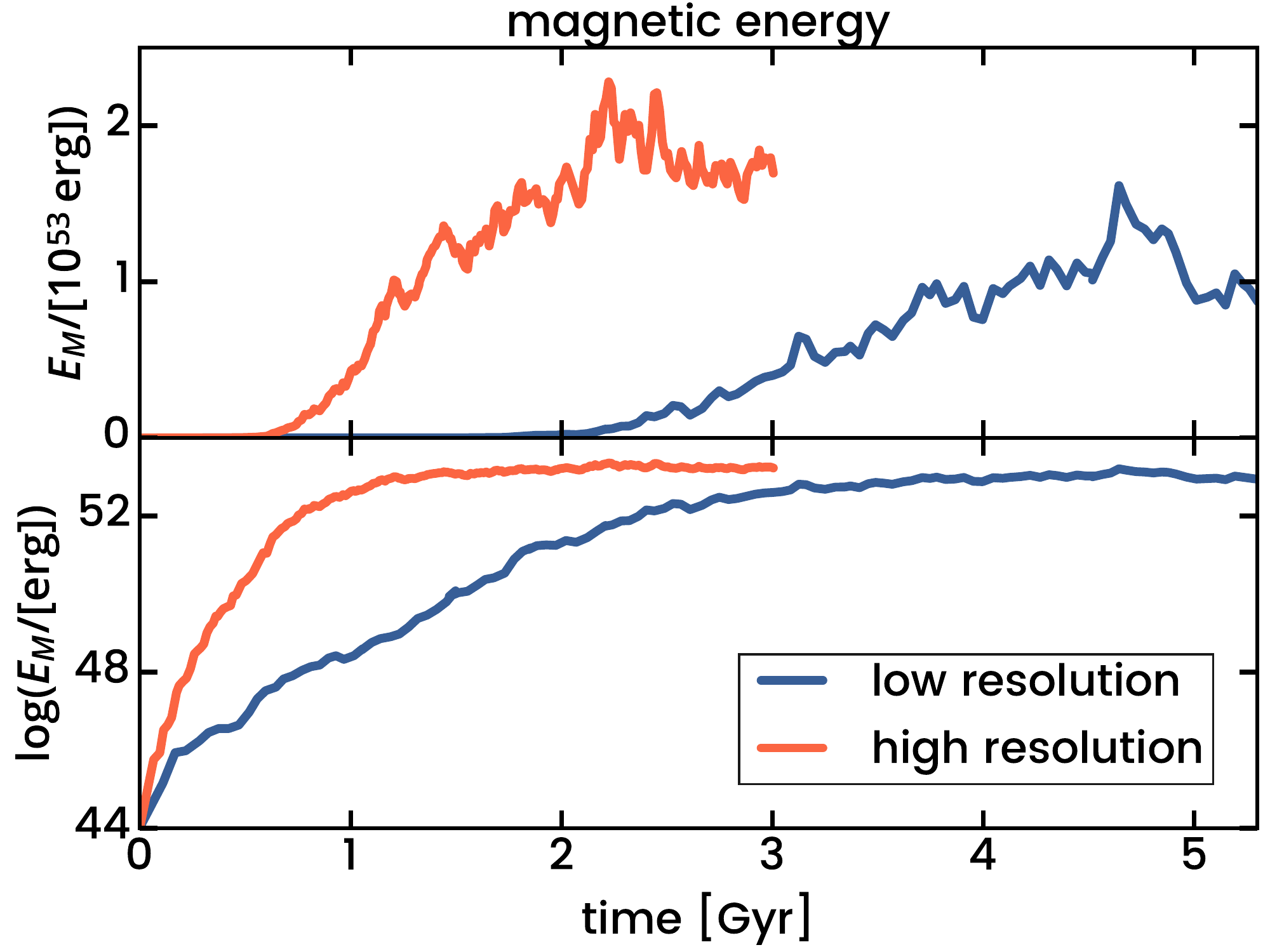}{Magnetic energy evolution in linear scale \textit{(top row)} and logarithmic scale \textit{(bottom row)} for $\Delta x = 18 \unit{pc}$ (low resolution) and $\Delta x = 9 \unit{pc}$ (high resolution). The high resolution run was run until 3 Gyr due to computational resource constraints.}{fig:emag}


\columnfigure{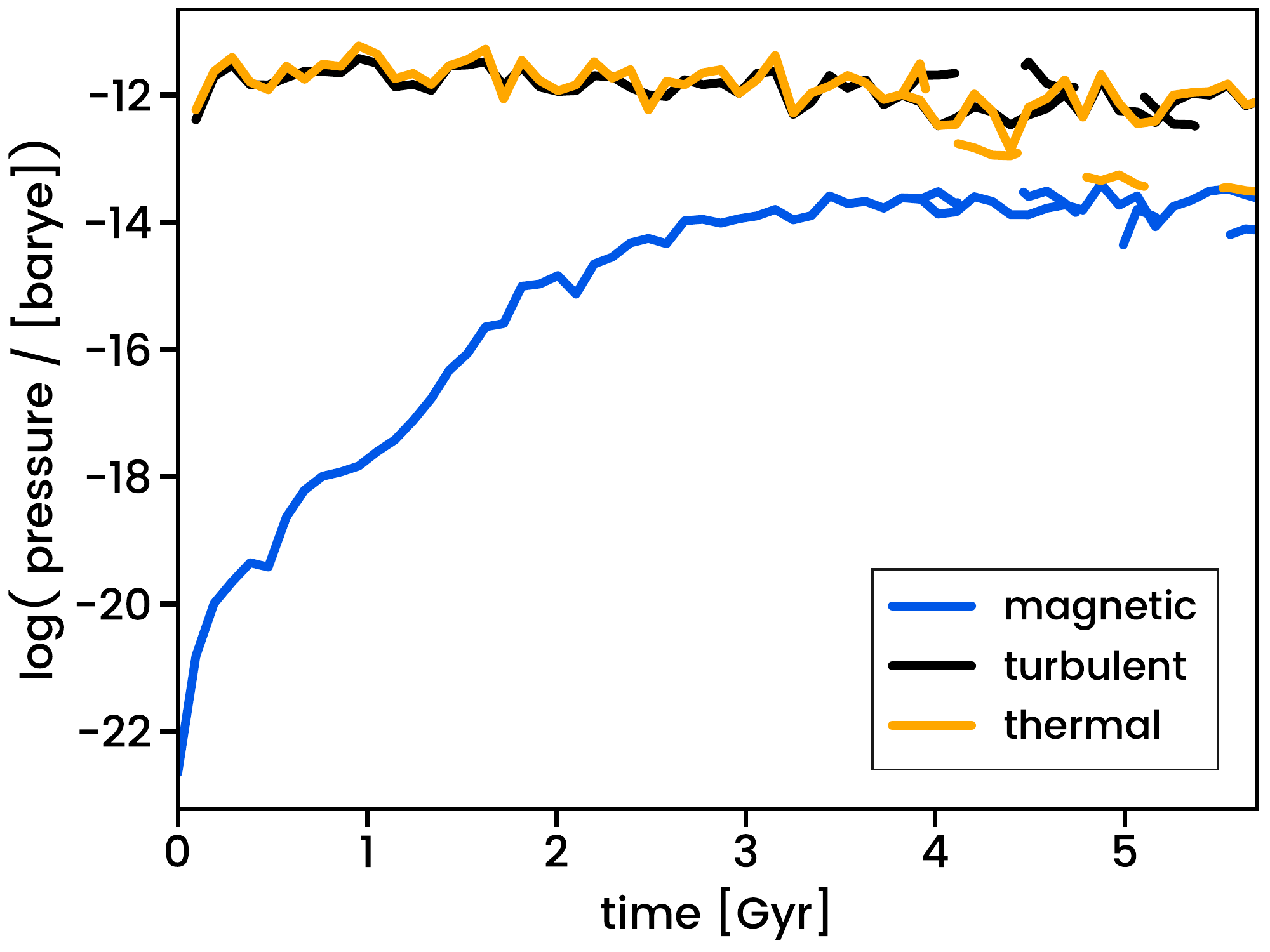}{Time evolution of thermal, turbulent and magnetic plasma pressure components in central 3 kpc cube for the feedback run \textit{(solid)} and the subsequent evolution when supernova feedback is suspended \textit{(dashed)}.}{fig:pressure}

In order to study the saturation level of the magnetic field, it is essential to measure the kinetic energy density of the turbulence. 
Defining and computing the turbulent energy in a realistic galactic environment  is however not as easy as in periodic boxes with forced or decaying turbulence. 
In our case, gas motions are dominated by the ordered galaxy rotation and quasi-random supernova explosions, 
both confined inside the stratified gravitational potential of the combined disk and dark halo system. 

As a proxy for turbulent energy, we use here the kinetic energy density of the velocity component perpendicular to the disk
\begin{equation}P_{\rm turb} \sim \frac 1 2 \rho u_z^2\end{equation}
since it is not affected by galactic rotation. 
The time evolution of the thermal, turbulent and magnetic energies, averaged in a cube of 3~kpc in the centre of the galaxy, is shown in \autoref{fig:pressure}. 
One can see that thermal and turbulent pressures are of the same order of magnitude, at approximately $10^{-12}$~erg/cc during the entire simulation. 
The magnetic pressure, on the other hand, defined as $P_{\rm mag} = \frac 1 {8\pi} B^2$, increases first exponentially, 
and then, at the onset of the non-linear dynamo phase, reaches only one per mil of the turbulent energy, corresponding also to a plasma 
$\beta = P_{\rm therm}/P_{\rm mag} \simeq 700$.
At the end of the simulation, when the dynamo is fully saturated, the average magnetic energy reaches a strength of $10^{-14}$~erg/cc, 
which is 2.5\% of the turbulent pressure, corresponding also to $\beta \simeq 40$.

\linefigure{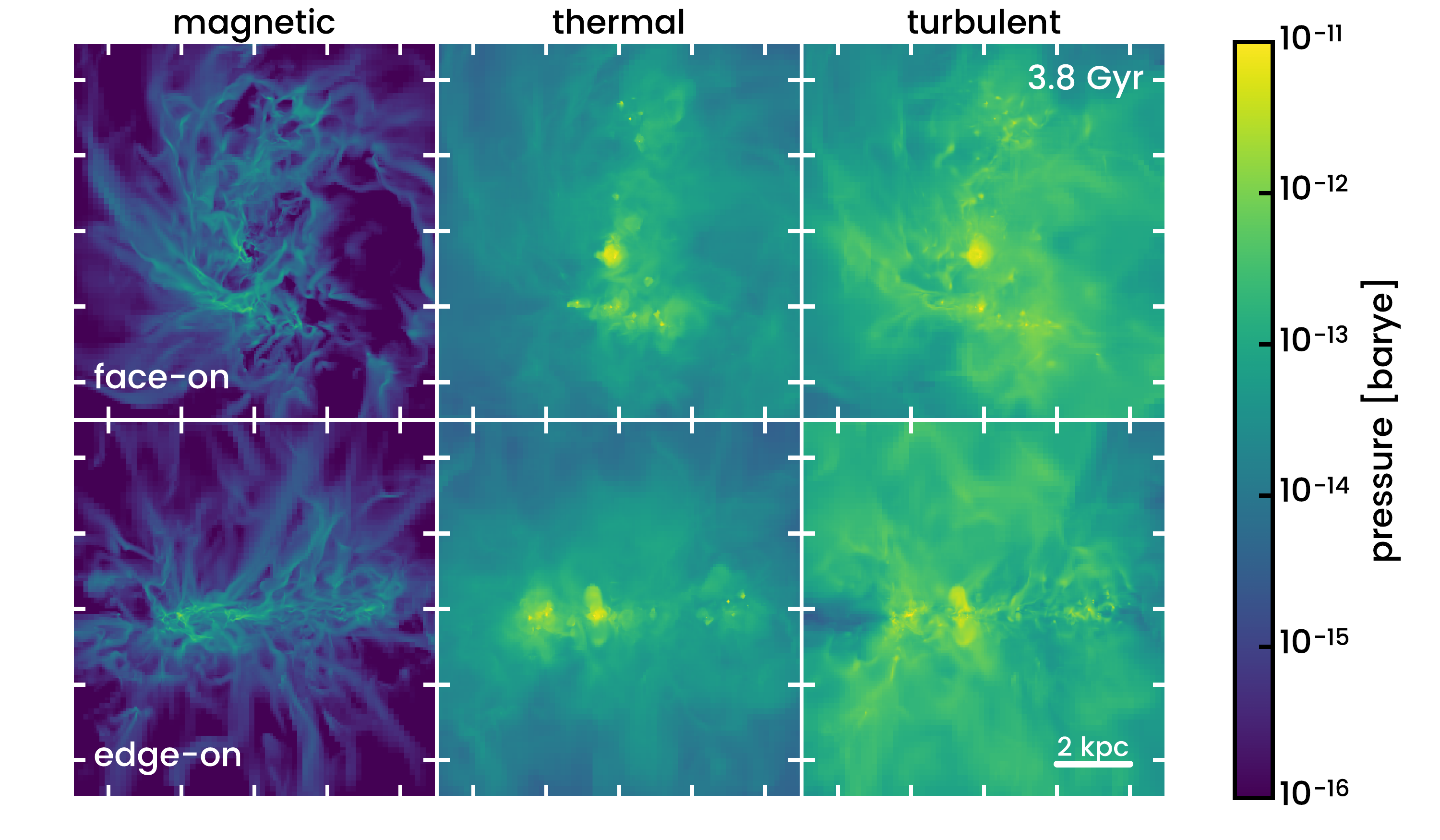}{Volume-weighted average magnetic \textit{(left)}, thermal \textit{(center)}, and turbulent \textit{(right)} pressure components along line of sight, looking face-on (top row) and edge-on (bottom row) in the $\Delta x = 18 \unit{pc}$ (lower resolution) simulation at the simulation time $t = 3.8 \unit{Gyr}$. Each panel covers 10 kpc, where every tick marks a distance of 2 kpc.}{fig:maps_fbk_pressure}

\autoref{fig:maps_fbk_pressure} shows volume-averaged line-of-sight projected maps of the different pressure components. 
One can clearly see that both thermal and turbulent pressure are relatively diffused and homogeneous, 
with only a few hot bubbles associated to supernova explosions appearing as distinctive features. 
The magnetic pressure map, on the other hand, is much more structured, with some striking filamentary features, resulting from a complex dynamo process. 
The magnetic pressure inside these filaments exceeds $10^{-12} \unit{barye}$, corresponding to a field strength $B > 1 \unit{\mu G}$.

\scalelinefigure{0.65}{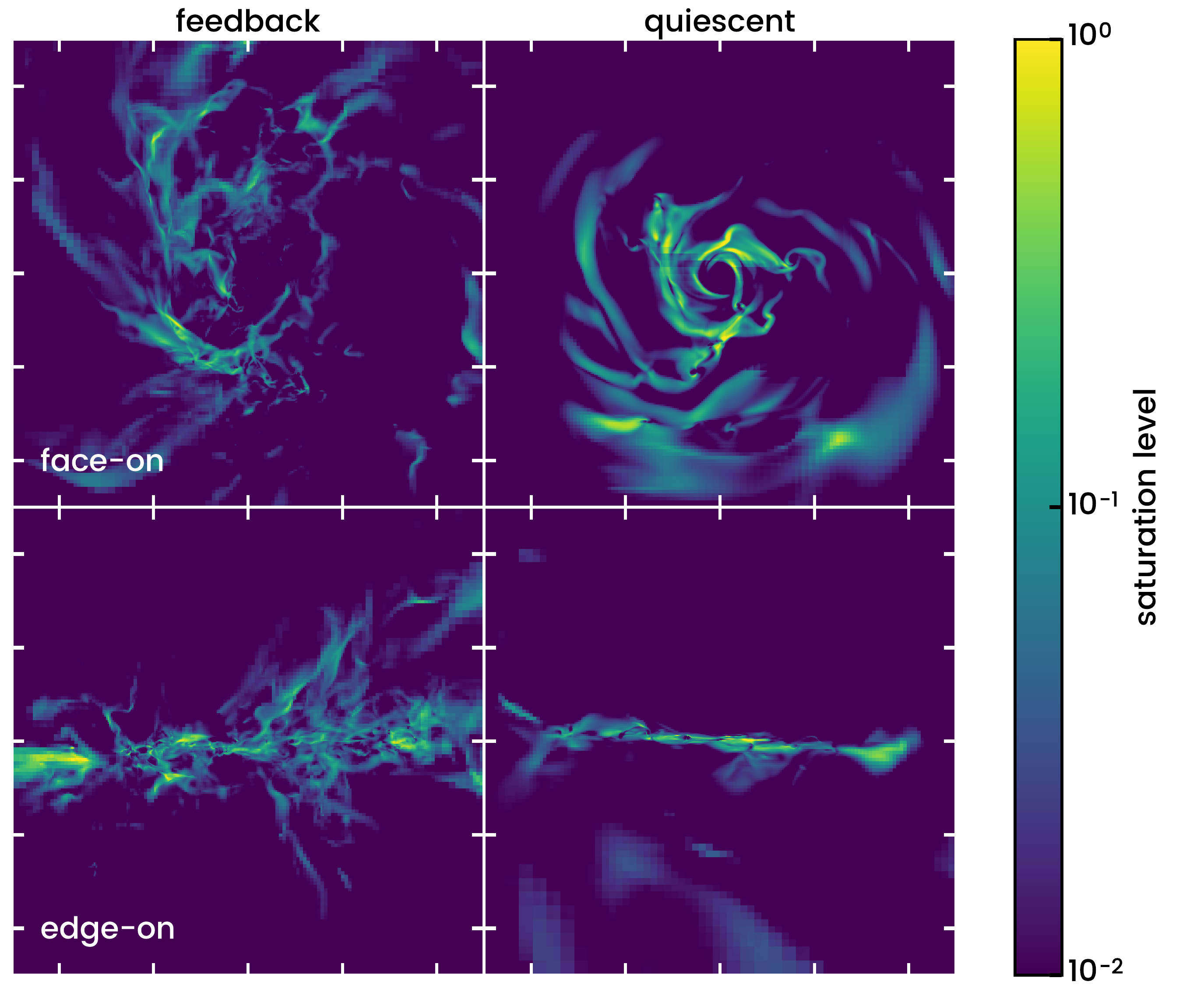}{Ratio of magnetic to turbulent pressure maps with feedback at time $t = 3.8 \unit{Gyr}$ \textit{(left)} and after feedback was switched off at the simulation time $t = 5.6 \unit{Gyr}$ \textit{(right)}, looking face-on \textit{(top row)} and edge-on \textit{(bottom row)}. Each panel covers 10 kpc, where every tick marks a distance of 2 kpc.}{fig:maps_sat}

We additionally plot the ratio of the magnetic to turbulent energy in \autoref{fig:maps_sat} (left panel) to evaluate to strength of the field at saturation.
Although the magnetic energy (or pressure) averaged over some volume is always (sometimes significantly) below equipartition (even after the saturation, as shown in \autoref{fig:pressure}), 
we can now see that the field  has actually reached equipartition in localised regions, usually associated with these strongly magnetised filaments.


We plot in \autoref{fig:spectra} the spectra of the magnetic and kinetic energy inside the galaxy. 
The kinetic energy spectrum stays roughly constant over time and exhibits a characteristic power-law behaviour
$E\subscript{kin} \propto k^\alpha$ with a best fit for $\alpha=-1.8$, 
which is between the theoretical values of $\alpha=-5/3$ for incompressible Kolmogorov turbulence 
and $\alpha=-2$ for highly compressible, shock-dominated Burgers turbulence. 
The magnetic energy spectrum develops also the characteristic shape predicted by Kazantsev's theory, with a power law on larger scales with index $1.5$, 
bottlenecked at small scales because of magnetic diffusion. 
During the kinematic phase, the magnetic energy power spectrum peaks at a length scale $\ell_{\rm peak} \simeq 500$~pc, 
and progressively shifts to larger scales until it reaches saturation, with a final peak length scale of $\ell_{\rm peak} \simeq 1 \unit{kpc}$.
This rather large spatial scale is a fundamental prediction of our theoretical calculations, 
and is justified by the very large spatial scale (around 10~kpc or more) at which the galactic fountains injects turbulent  energy into the magnetic dynamo.

\columnfigure{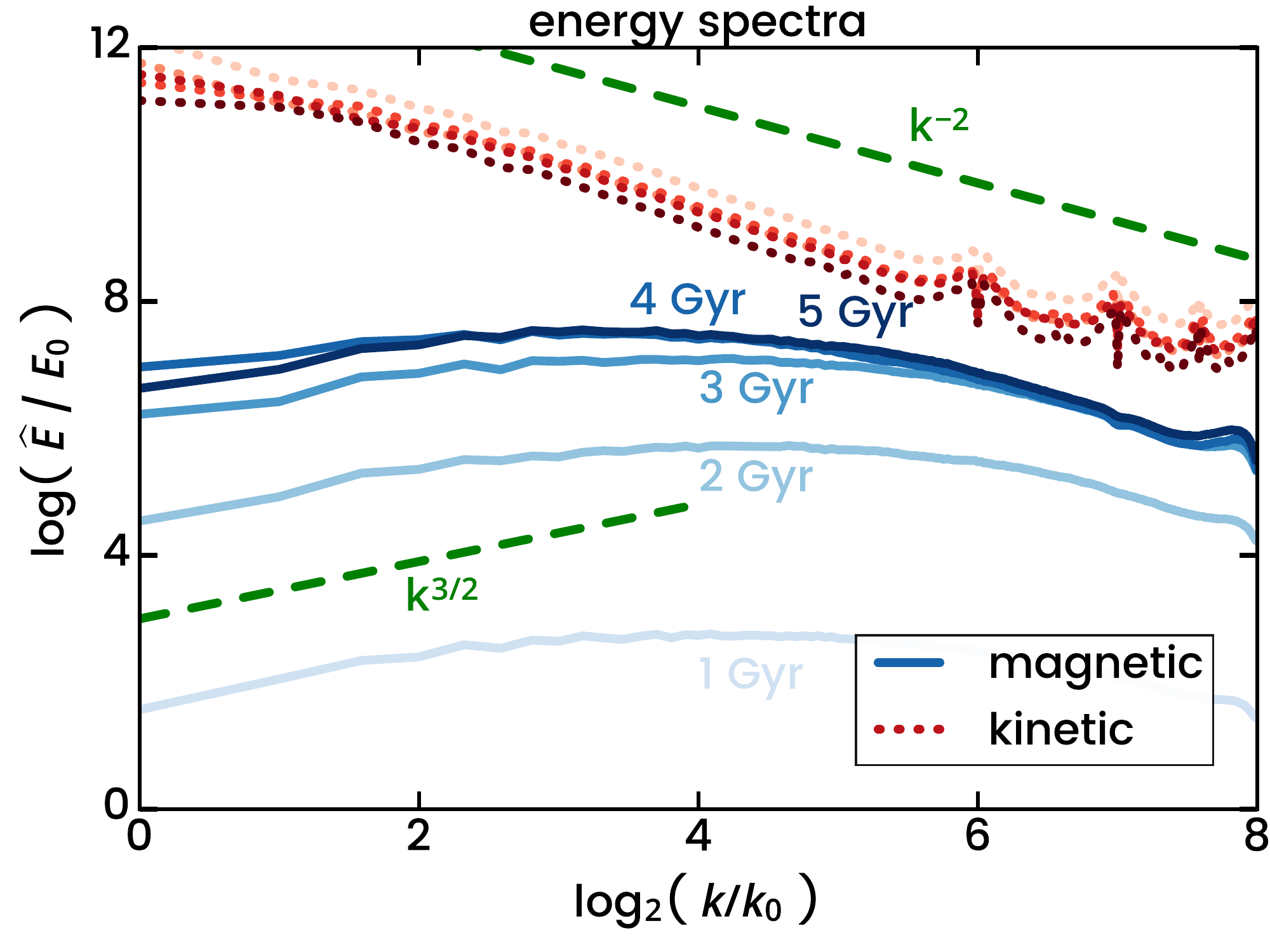}{\textit{Left:} Spectra of kinetic and magnetic energies in central $512^3$ cube box at grid resolution, resulting in cube size of 9.4 kpc. The normalization factor $E_0$ is the initial magnetic energy integrated over the whole spectrum. \textit{Right:} Comparison of the same spectra at saturation to the kinetic and magnetic energy spectra in the high-resolution run with $1024^3$ points.}{fig:spectra}

Interestingly, our saturated dynamo being strongly coupled to a large scale galactic fountain, we are in a position to provide strong observational predictions on the strength of the magnetic field
around high-redshift galaxies. Indeed,  magnetic fields in the circumgalactic medium can be detected by the Faraday rotation of polarised emission from background sources, 
with impact parameters as large as the virial radius of the parent halo \citep{M-2008Natur.454..302B}. 
We plot in \autoref{fig:sphrad} the radial profiles of the mean gas density, 
the gas metallicity and the magnetic field strength averaged in spherical shells around the galaxy out to the virial radius at 30~kpc.
We can see that the galactic fountain is polluting the IGM around the galaxy, with a metallicity of $Z \approx 0.2 Z_{\odot}$ and the magnetic field strength declining from $\mu$G to a tenth of a nG, and the gas density from $1 \unit{H/cc}$ to $\sim 10^{-4}$~H/cc. 

\columnfigure{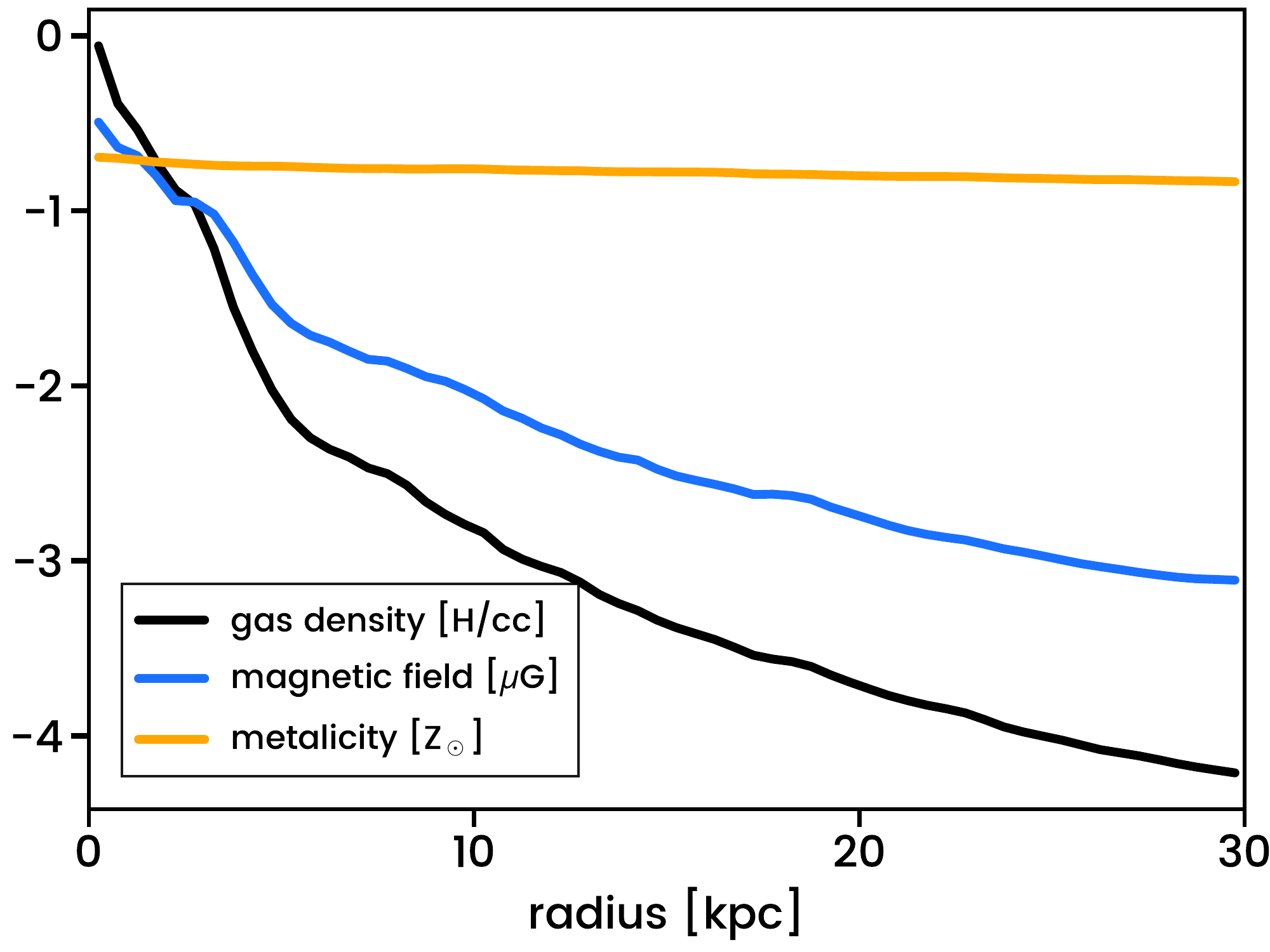}{Radial profiles of average gas density, metalicity, and magnetic field strength in spherical shells around the galaxy out to large radii from the galaxy with strong feedback.}{fig:sphrad}

\subsection{Effect of Resolution}


The small-scale dynamo growth rate increases with the effective Reynolds number, which in our case is set by the numerical diffusion of our Godunov scheme. 
In order to study this numerical effect, we increase the resolution by one level to reduce both magnetic diffusion and kinematic viscosity, 
so that both effective Reynolds numbers (magnetic and kinematic) become larger. 
Due to limited computational resources, we could only run this simulation for 3~Gyr. 
However, since the dynamo operates more than twice as fast at this increased resolution, it reached saturation before the lower resolution run.

The magnetic energy evolution for this high-resolution run is plotted alongside the low resolution evolution in \autoref{fig:emag}. 
The behaviour is qualitatively the same for both. However, the growth rate in the kinematic phase is larger with $\Gamma \simeq 9.4 \unit{Gyr}^{-1}$ in the high resolution case. 
Consequently, the onset of the non-linear dynamo phase happens earlier. 
The overall magnetic energy level at saturation is slightly higher than in the low-resolution run, 
so that the magnetic field saturates at $5 \%$ of turbulent energy density, corresponding also to a final plasma $\beta \simeq 20$.

We plot in \autoref{fig:spectra_res} the kinetic and magnetic energy spectra from the last snapshot of the high resolution run, compared to the same spectra at saturation for the low resolution run. 
The high-resolution kinetic spectrum has the same turbulent slope as the low-resolution one, with however a slightly higher amplitude, owing to the slightly more efficient star formation efficiency and associated supernovae feedback. The magnetic spectra for the high resolution run also agrees well with the low resolution one at large scale, with however slightly more power on small scales, due to the decrease of the effective dissipation length. This results in a slight shift of the peak scale from $\ell_{\rm peak}=1$~kpc to  $\ell_{\rm peak}=0.7$~kpc.

\columnfigure{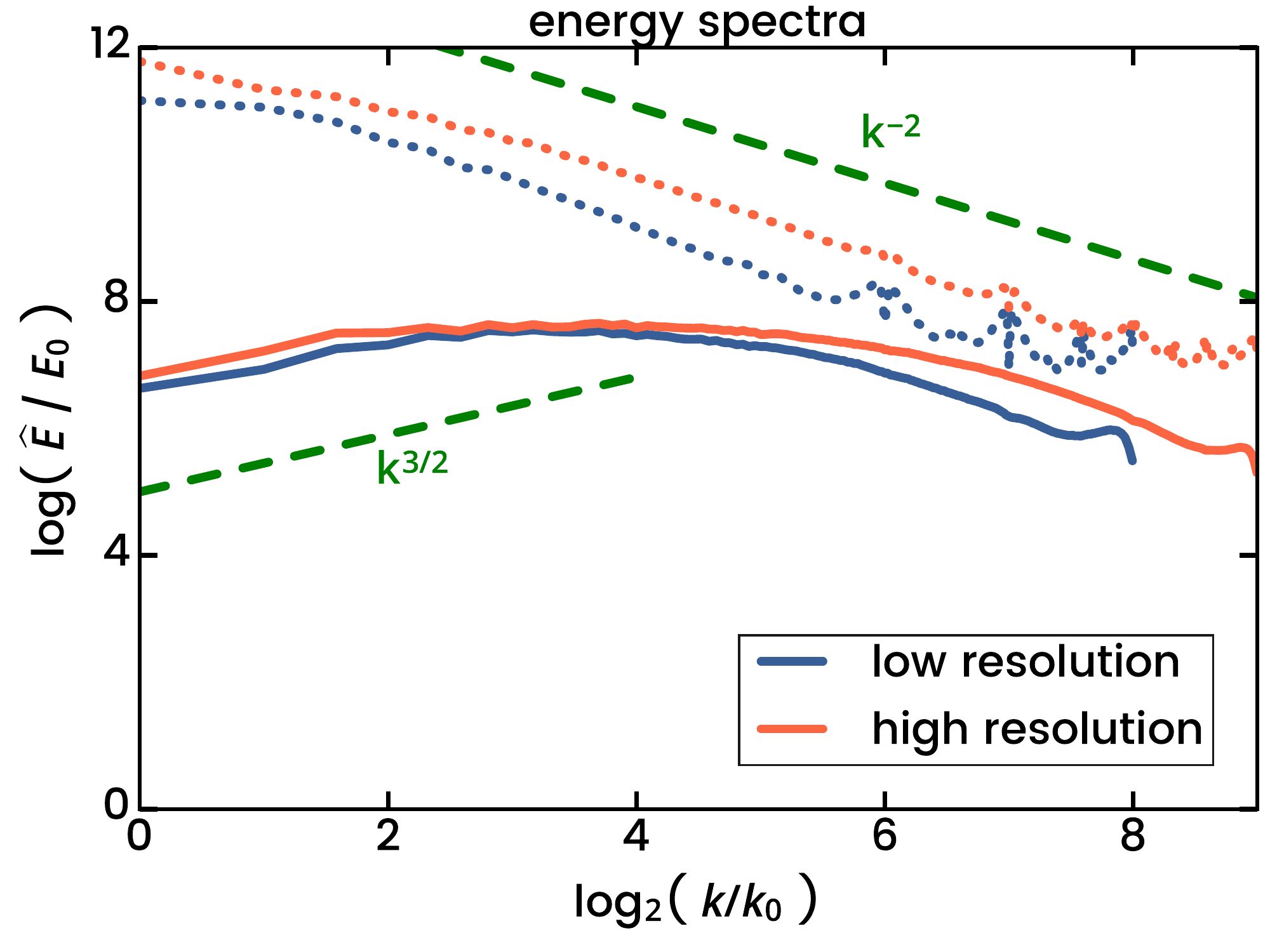}{\textit{Left:} Spectra of kinetic and magnetic energies in central $512^3$ cube box at grid resolution, resulting in cube size of 9.4 kpc. The normalization factor $E_0$ is the initial magnetic energy integrated over the whole spectrum. \textit{Right:} Comparison of the same spectra at saturation to the kinetic and magnetic energy spectra in the high-resolution run with $1024^3$ points.}{fig:spectra_res}

\subsection{Transition to Quiescence}

\triplefigure{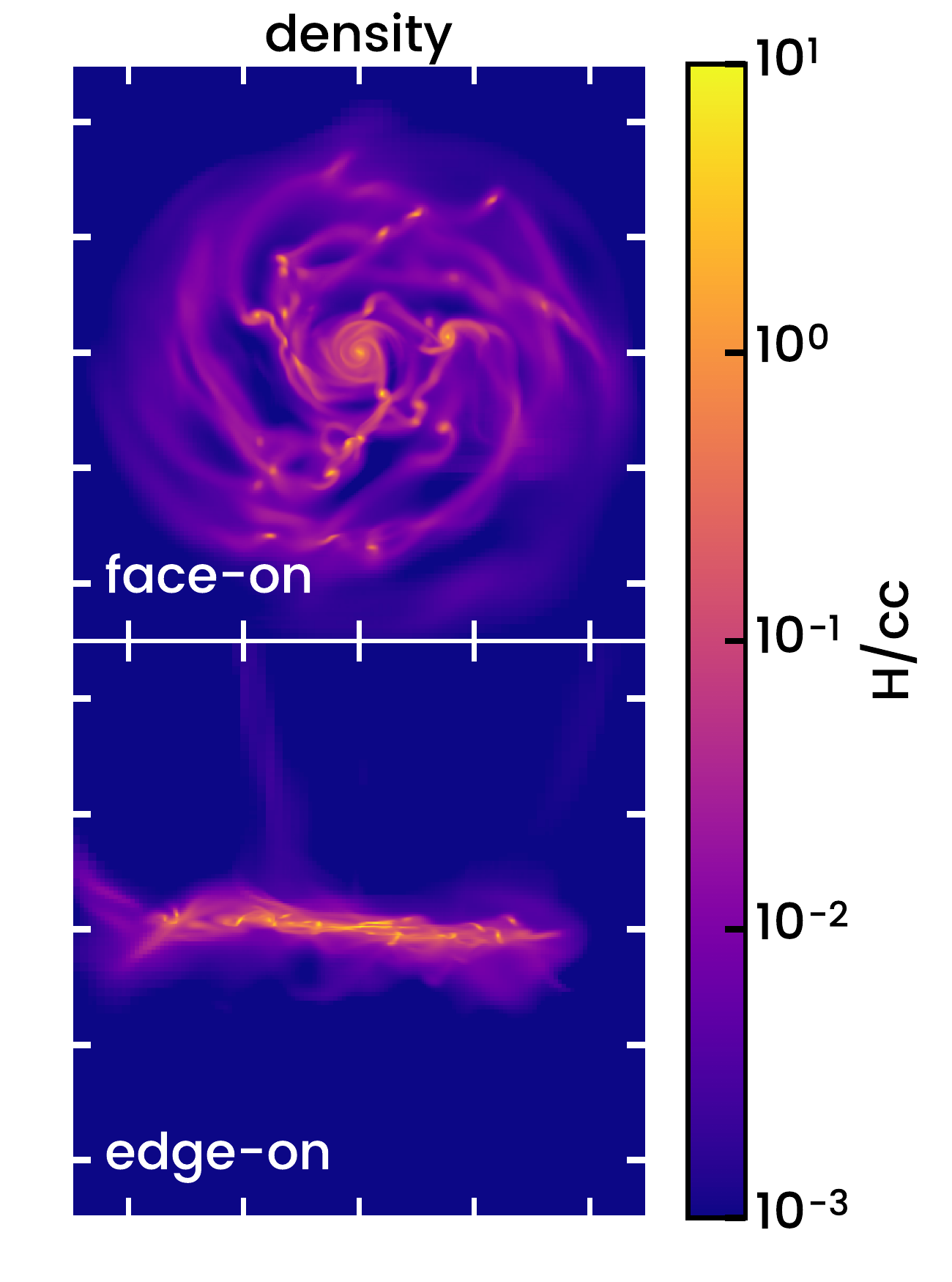}{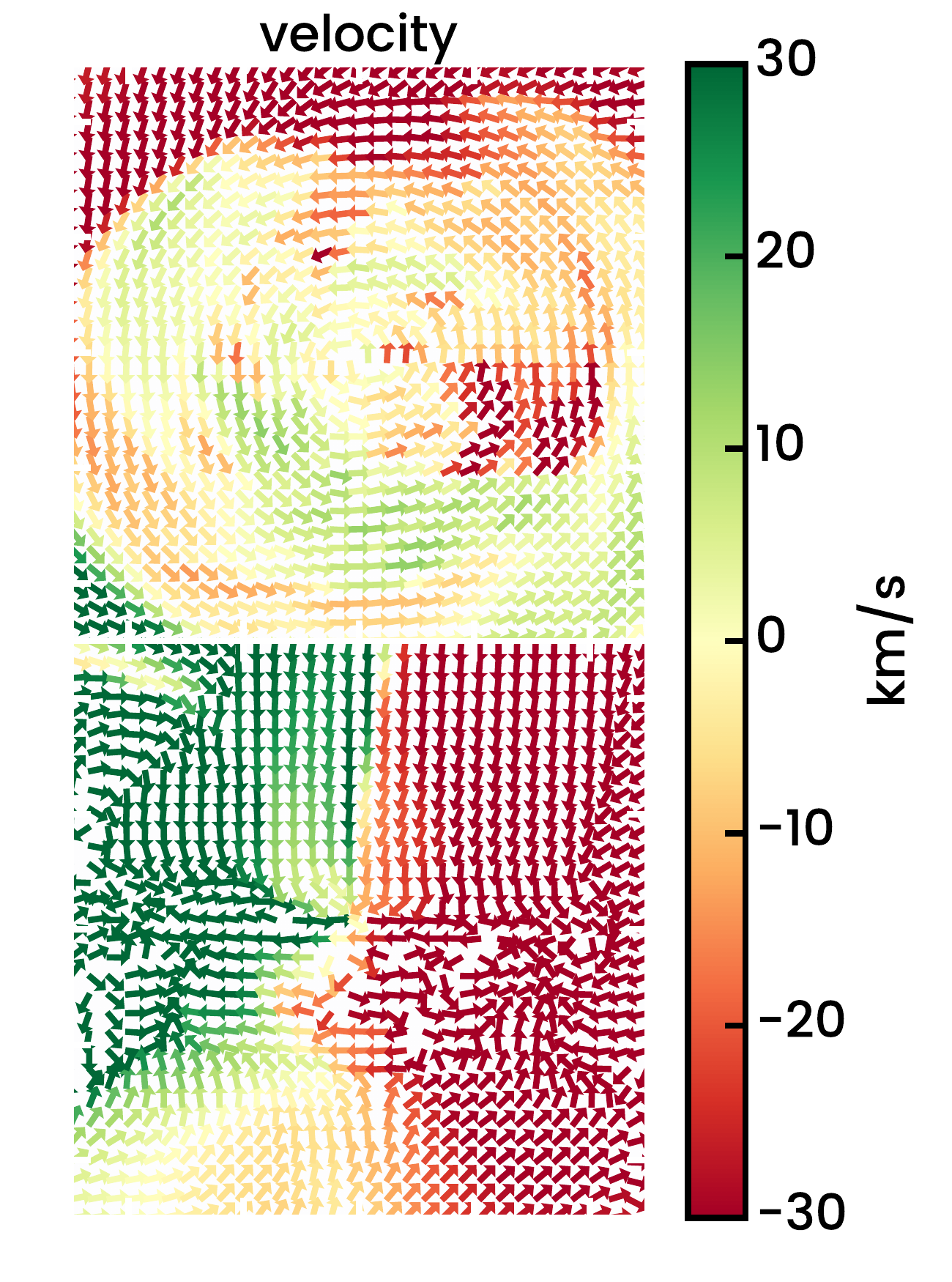}{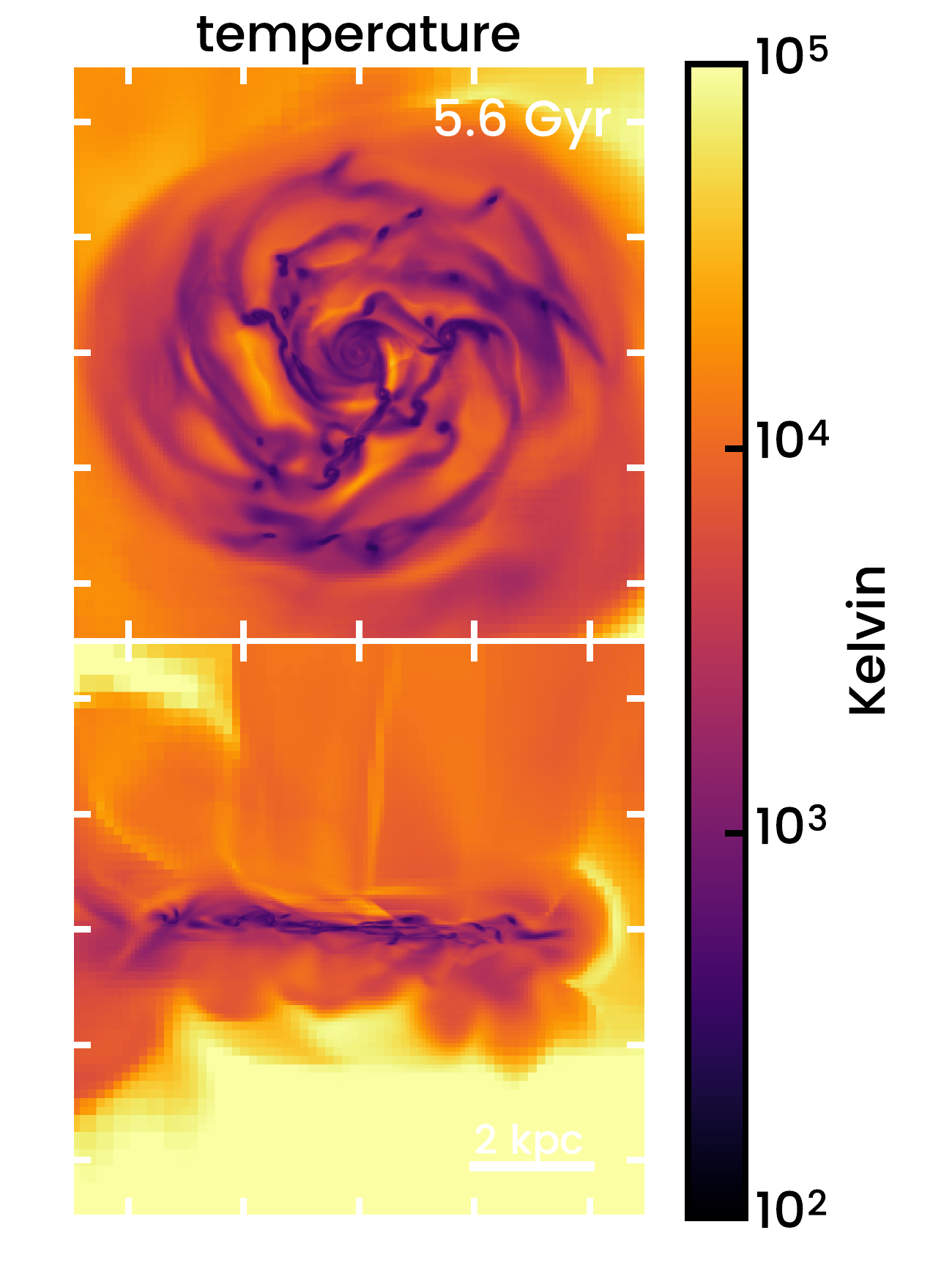}{Volume-averaged line-of-sight projections for different plasma properties looking face-on \textit{(top row)} and edge-on \textit{(bottom row)} in the central 10 kpc cube in the $\Delta x = 18 \unit{pc}$ (lower resolution) simulation at the simulation time $t = 5.6 \unit{Gyr}$, or 1.6~Gyr after feedback was switched off. \textit{Left:} density \textit{Center:} velocity \textit{Right:} temperature.}{fig:maps_qsc}

We have seen in the previous section that in the active, early phase of its life, dominated by stellar feedback, 
a galaxy with a fully developed turbulent fountain can power a small-scale dynamo that efficiently amplifies an initially weal seed magnetic fields to saturation. 
Naturally, one might wonder what would happen to this sub-equipartition field, if feedback becomes weaker. 
Indeed, present-day galaxies like the Milky Way have thin and quiescent disks, with a modest level of turbulence and a small kinetic energy injection scale around 100~pc, 
traditionally associated to the thickness of the gas disk or to local supernova super-bubbles \citep{1992ApJ...391..188F}. 
For that purpose, we re-run the simulation from our snapshot at 4~Gyr but with supernova feedback turned off and let it evolve for a couple of Gyr.

In \autoref{fig:maps_qsc}, we show images of our galaxy at time 5.6 Gyr, 1.6~Gyr after feedback has been switched off. 
Without feedback to drive the turbulence anymore, the galaxy has entered a quiescent phase where the gas has cooled down 
and collapsed into a thin, although clumpy, rotationally supported disk,
with a clear spiral structure. The velocity field is dominated by a strong rotational component, anti-clockwise in this image, which means that the spirals are trailing.

The evolution of the thermal, turbulent and magnetic pressure components is shown with dashed lines in \autoref{fig:pressure}. 
The average magnetic energy density decreases only slightly when the disk enters this new phase and remains at a level of $10^{-14}$ erg/cc. 
At the same time, we can see a clear drop in thermal energy density and, less pronounced, in the turbulent component. 
The latter, however, since we use the vertical gas velocity component as a proxy for turbulence, is still contaminated by a vertically collapsing flow.

\linefigure{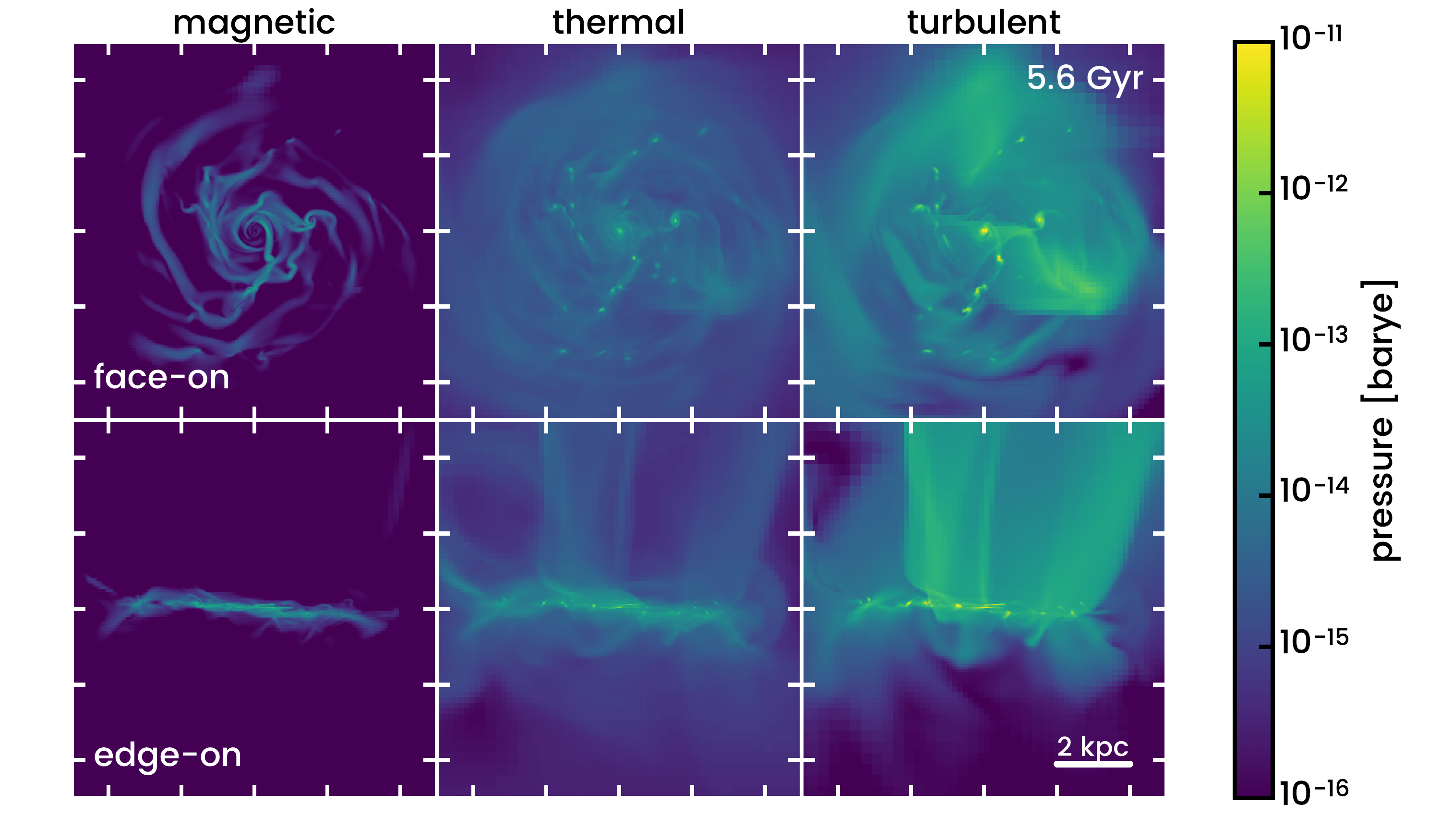}{Volume-weighted average magnetic \textit{(left)}, thermal \textit{(center)}, and turbulent \textit{(right)} pressure components along line of sight, looking face-on \textit{(top row)} and edge-on \textit{(bottom row)} in the $\Delta x = 18 \unit{pc}$ (lower resolution) simulation after feedback was switched off at the simulation time $t = 5.6 \unit{Gyr}$. Each panel covers 10 kpc, where every tick marks a distance of 2 kpc.}{fig:maps_qsc_pressure}

To characterise the magnetic field strength in our thin disk galaxy even further, we compare the magnetic pressure to the gas thermal and turbulent energy densities side-by-side, 
by plotting line-of-sight projections of all three pressure components in \autoref{fig:maps_qsc_pressure}. 
The contamination of our turbulent energy proxy by the vertical collapse is clearly visible from the in-falling gas above the disk.
While they all have significantly decayed inside the galactic corona (except for the vertical component of the velocity field), 
they all remain strong inside the disk. Notably, the energy density of the magnetic field is now in equipartition with the other two energies inside the galactic arms, 
exceeding $10^{-12} \unit{barye}$, which amounts to field strengths greater or around $1 \unit{\mu G}$. 
We have also plotted the ratio of the magnetic to turbulent pressure in the quiescent case in the right panel of \autoref{fig:maps_sat}. 
The ratio reaches unity inside the arms, meaning that the magnetic field is in fact {\it locally} 
in equipartition with turbulence and thermal pressure in the dense galactic arms of our thin disk.

In order to characterise the magnetic field topology, 
we plot in the top panel of \autoref{fig:vert_symmetry} the toroidal, radial and vertical components of the field, as a function of the disk height,
averaged within cylindrical shells of different sizes, parallel to the disk plane. 
On can see that after the collapse of the fountain into a thin disk, the vertical component of the field cancelled almost entirely, leaving only 
a dominant toroidal component, with a clear even-symmetry with respect to the disc plane, and a very weak average radial component.
We argue here that this quadrupole symmetry is a natural consequence of a random magnetic field collapsing into the midplane, 
with a cancellation of odd-symmetric modes and a strengthening of even-symmetric ones during the formation of the thin disk. 
On the other hand, one also clearly sees sign reversals in the average toroidal component, so that its average over the entire disk remains very small.

\columnfigure{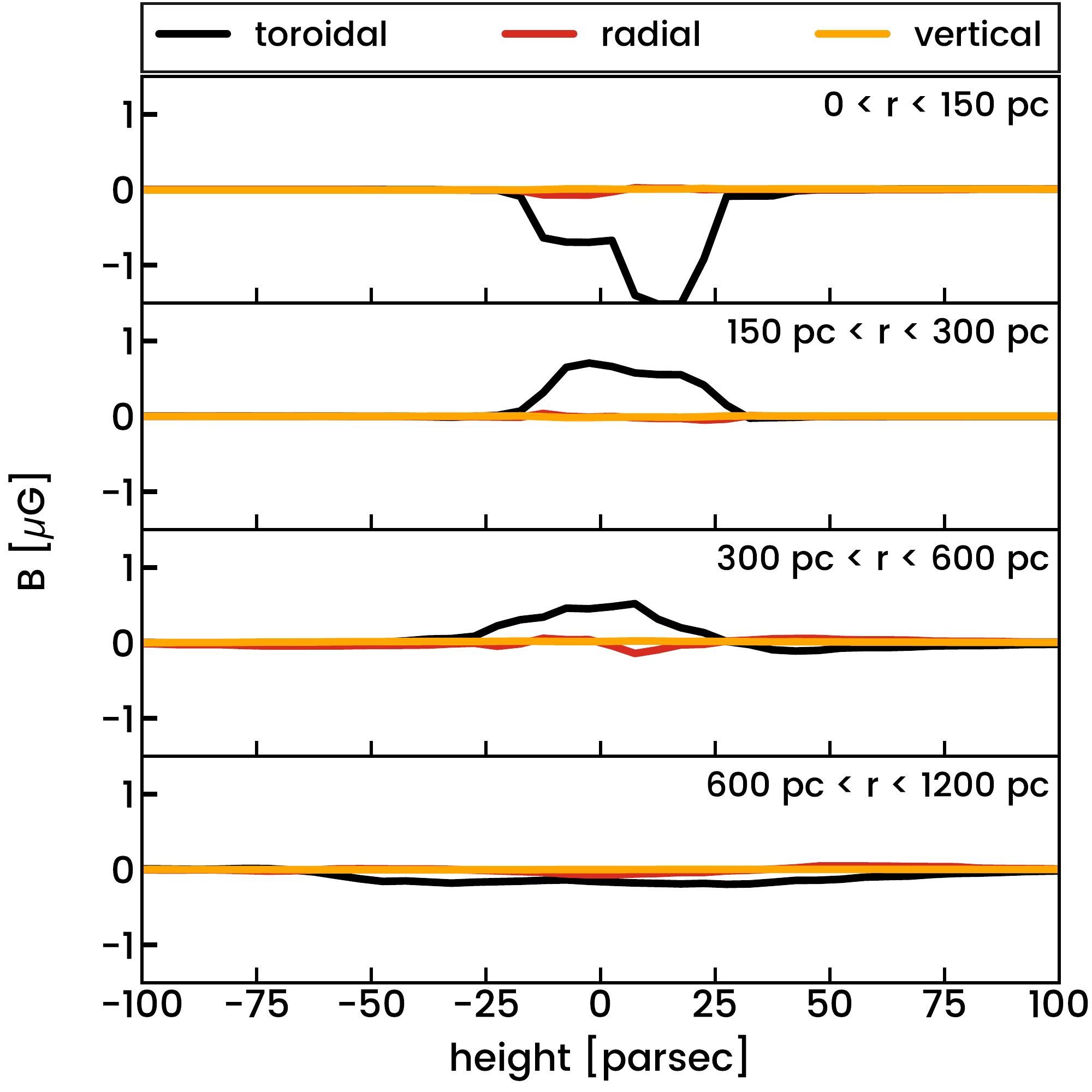}{Average toroidal, radial and vertical magnetic field components as a function of the height relative to the mid-plane of the quiescent disk in cylindrical shells of different sizes.}{fig:vert_symmetry}

In \autoref{fig:radialprof}, we plot the three magnetic field components as a function of radius, averaged over very thin cylindrical shells of thickness $H = 40$~pc.
Inside the innermost region up to 1 kpc, the magnetic field is dominated by the toroidal field component with field strengths up to $1 \unit{\mu G}$. 
One can also see in the radial profile significant sign reversals of the toroidal component. 
These sign reversals are a clear relic of the small scale dynamo in the feedback-dominated corona prior to the collapse, and are also tightly connected to the spiral pattern.
In the lower panel of \autoref{fig:radialprof}, we show the standard deviation of the three field components, averaged within the same small cylinders.
This quantity is a measure of the field strength at small scale, and reaches 1~$\unit{\mu G}$ for the toroidal component, roughly in equipartition with the mean field.
The standard deviation for the vertical component is almost zero, while the small scale radial component has a field strength around 0.3-0.4~$\unit{\mu G}$.

\columnfigure{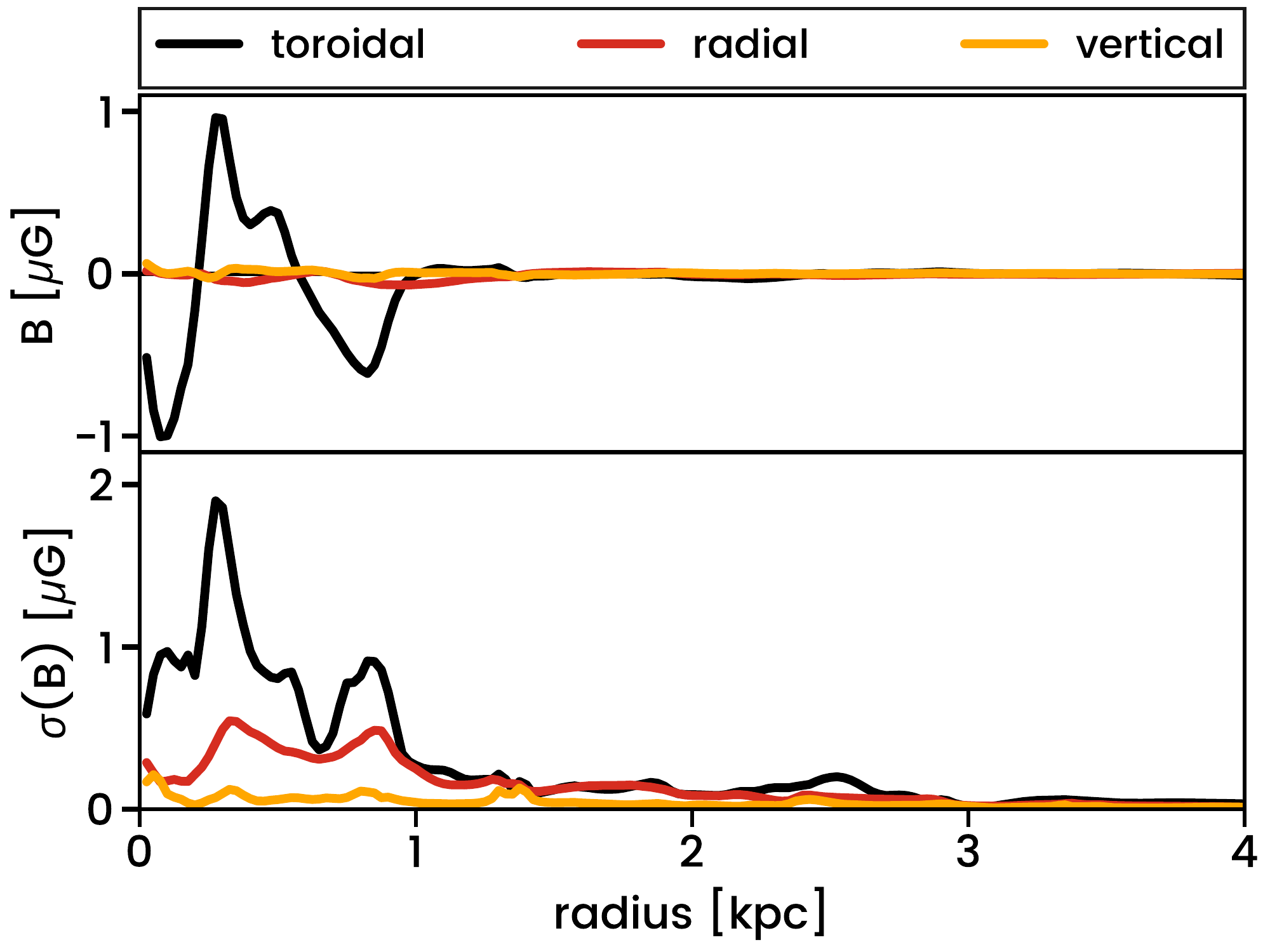}{Averaged profiles of magnetic field \emph{(top)} and its standard deviation \emph{(bottom)} for the toroidal, 
radial and vertical field components along the radius in cylindrical shells around the quiescent galaxy.}{fig:radialprof}

To quantify and illustrate the relative strength of the radial and toroidal components better, we show the magnetic field line directions ${B}_{\bot}$ perpendicular to the line of sight in a face-on view of the average magnetic field in \autoref{fig:pitchangle}, together with a histogram of the pitch angle
\begin{equation} p_B = \fn{arctan}{\frac{B_r}{B_t}}. \end{equation}
which is a common observable when measuring galactic magnetic fields. The magnetic field is strongest inside the trailing main arms which also coincide with the dense gas arms. Its field lines are generally aligned with the arm structure in the arms and otherwise mostly pointing along the toroidal direction (sometimes parallel, sometimes anti-parallel). We find that the pitch angles are symmetrically distributed around a mean pitch angle of $\langle p\subscript{B} \rangle = -12.4 \unit{^\circ}$ with a standard deviation of $\sigma(p\subscript{B}) = 35.1 \unit{^\circ}$. Note that these negative pitch angles are a clear indicator of magnetic field alignment with the spiral structure of a trailing spiral galaxy and change sign because of the field reversals.

\towerfigure{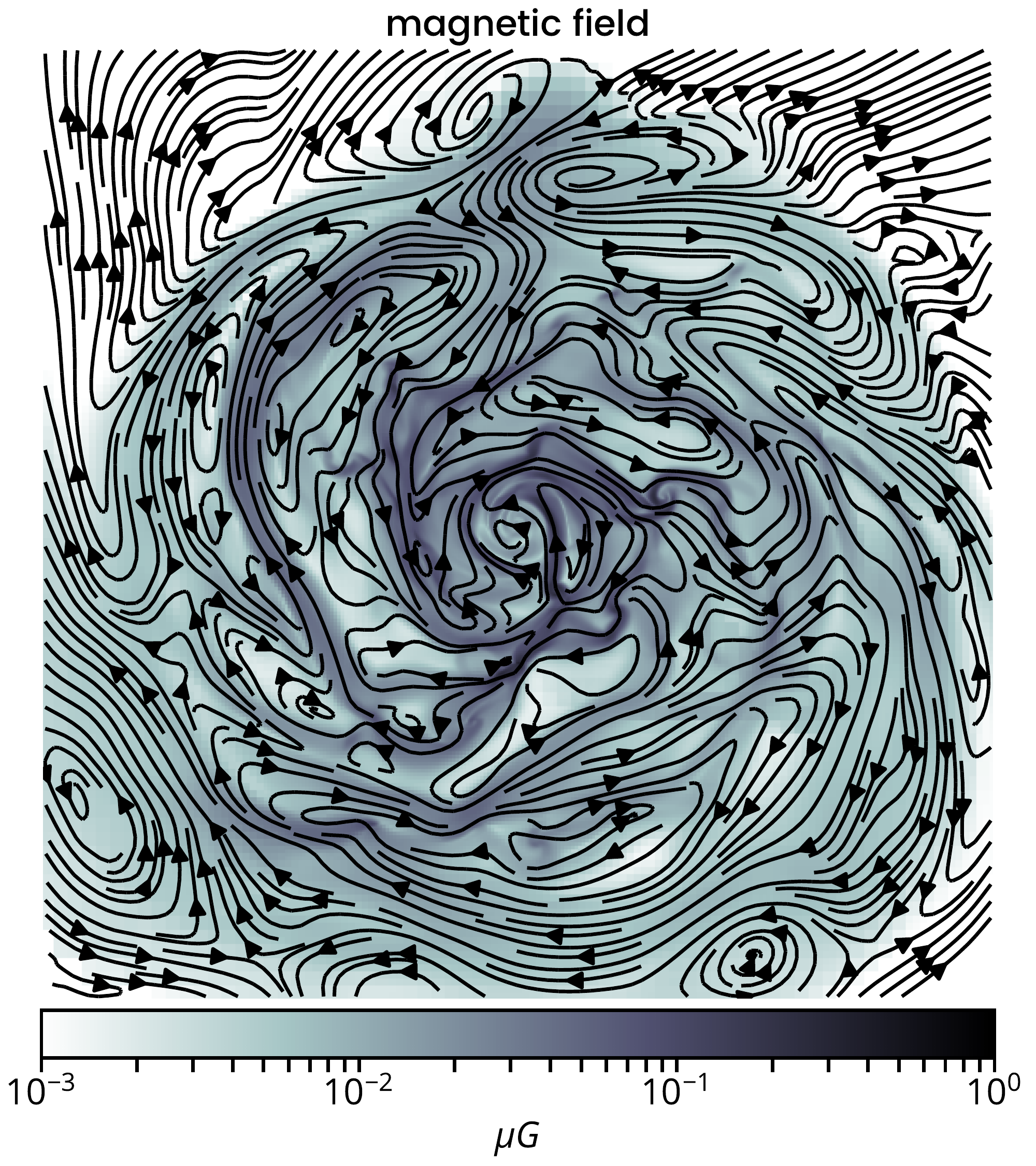}{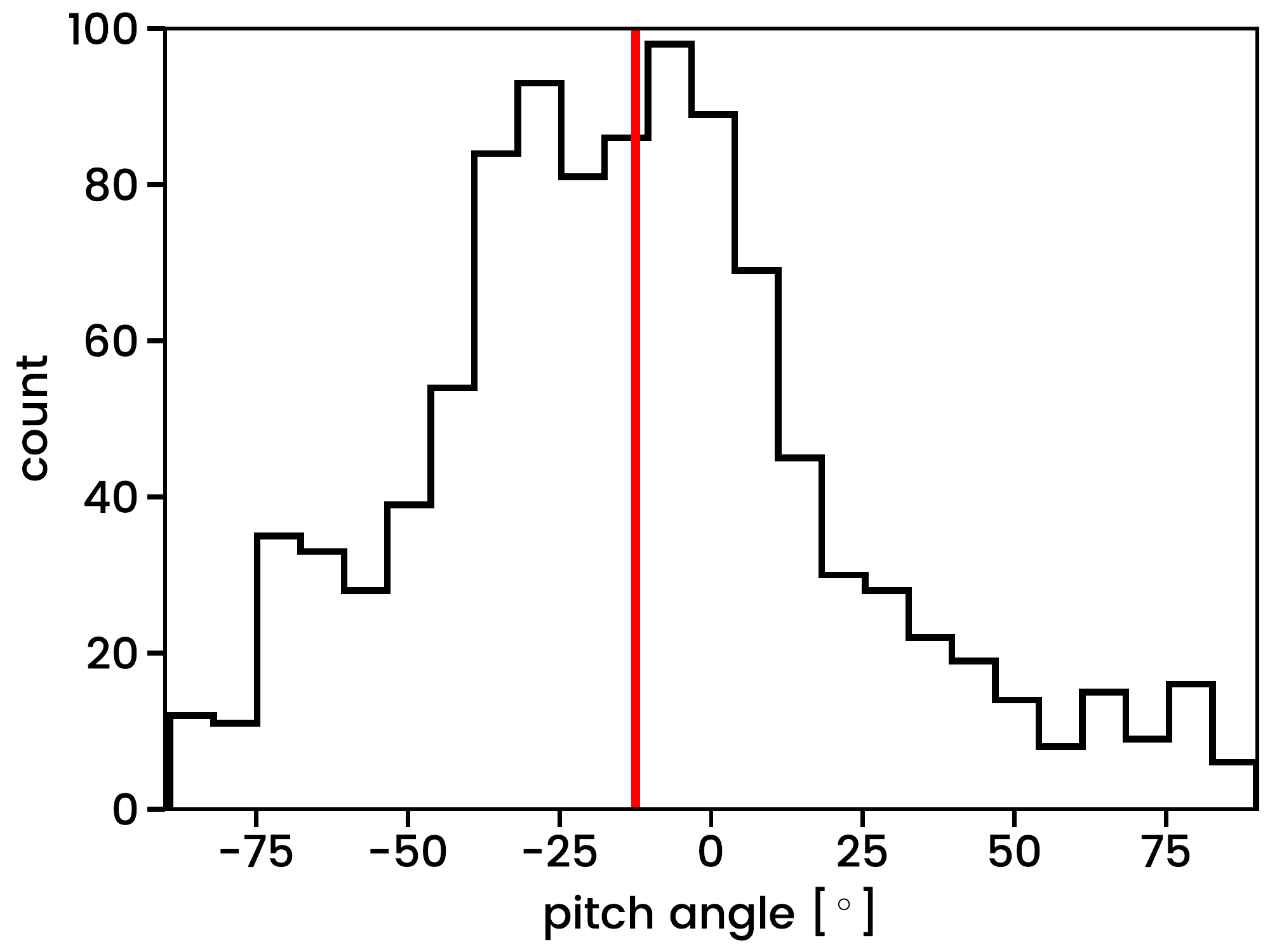}{Magnetic field line directions on top of a line-of-sight averaged magnetic field map in a face-on view \emph{(top panel)} and the count histogram of pitch angles in above image \emph{(bottom panel)}. The red line marks the mean pitch angle $\langle p\subscript{B} \rangle = -12.4 \unit{^\circ}$ with a standard deviation of $35.1 \unit{^\circ}$.}{fig:pitchangle}

\section{Discussion}
\label{chap:discussion}

We have set up magnetohydrodynamical simulations of an isolated cooling dwarf halo which hosts a galaxy with strong turbulence driven by supernova feedback. The gas component features high Mach number flows with up to $M \approx 10$ and a kinetic energy spectrum with a characteristic slope between incompressible Kolmogorov and highly compressible shock-driven Burgers turbulence. Just as in our previous work \citep{2016MNRAS.457.1722R}, the magnetic seed field is quickly amplified by a small-scale dynamo. The magnetic energy spectrum has the typical bottlenecked $3/2$-slope on large scales which peaks near the resistive scale and falls off on smaller scales as predicted by dynamo theory \citep{M-1992ApJ...396..606K}. The global magnetic field time evolution  in our experiment can be divided into three distinct phases:
\begin{enumerate}
\item exponential growth (kinematic phase)
\item constant growth (non-linear phase)
\item zero growth (saturation)
\end{enumerate}
During the exponential growth phase, the e-folding times are $\sim 300 \unit{Myr}$ with lower resolution and $\sim 100 \unit{Myr}$ with higher resolution. Since the dynamo growth rate increases with the Reynolds numbers of the flow, it could potentially become very large if we were not constrained by our limited computational resources. Estimating the Reynolds numbers in a simulation with numerical viscosity and diffusivity can be very difficult when not using explicit terms in the system of equations and when complicated flows are involved. As a lower limit, we can take the numerical diffusion of a first-order method which is given by
\begin{equation}\mathrm{{Re}_{N}} = 2 \frac L {\Delta x}\end{equation}
with the typical length scale of the system $L$ and the spatial resolution $\Delta x$. This would give values for $\mathrm{{Re}_{N}} = 100$ in the lower resolution case  and 200 for the higher resolution in our simulations. Generally, in this kind of simulations, both the kinematic as well as the magnetic Reynolds number will range between a few $100$ and $1000$. We can compare this to a typical ISM with an estimate for the ISM viscosity \citep{1941ApJ....93..369S}
\begin{equation}\fn{Re} = \sqrt{3\gamma} M \frac L \lambda\end{equation}
where $\lambda$ is the mean free path of a hydrogen atom and $M$ is the Mach number. At a density of $1 \unit{H/cc}$ and a Mach number of 5 the Reynolds number will be $\fn{Re} = 10^7$. The magnetic Reynolds number computed from the Spitzer resistivity formula at $T = 10^4 \unit{K}$ is even larger with $\fn{Rm} = 10^{21}$, yielding the magnetic Prandtl number $\fn{Pm} = 10^{14}$. \cite{2015PhRvE..92b3010S} derive a scaling for the small-scale dynamo growth rate in the case of large $\mathrm{Pm}$ and supersonic turbulence as
\begin{equation}\Gamma \propto \mathrm{Re}^{1/2}\end{equation}
meaning that we can extrapolate the growth rate we measured in our simulation to expected ISM Reynolds numbers obtaining the growth rate with a realistic galactic small-scale dynamo of $\Gamma = 1000 \unit{Gyr^{-1}}$ and an e-folding time of just $\tau = 1 \unit{Myr}$ respectively.

At the time of saturation, the magnetic pressure to turbulent pressure ratio is between 2.6\% in the lower resolution case and 5\% for the higher resolution. All of this is in agreement with the theoretical results of \cite{2015PhRvE..92b3010S} who predict the same three-phase evolution and give expected saturation levels. Although those numbers apply only in the limiting cases of very small or very large magnetic Prandtl numbers, they are still consistent with our result lying in between due to $\unit{Pm} \sim 1$ from both Reynolds numbers being caused numerical scheme. The saturation values are also in agreement with prior numerical studies, such as \cite{2011PhRvL.107k4504F} who find 2\% with $\unit{Pm} = 2$ and solenoidal forcing at Mach 10 and \cite{2014ApJ...797L..19F} who find 3\% for $\unit{Pm} = 2$ and 5\% for $\unit{Pm} = 5$ (both at Mach 11). The same holds for \cite{2016MNRAS.461.1260T} who report saturation ratios between 2\% and 4\% and also point out a trend of an increased ratio in the case of higher resolutions. The agreement of our results with the simulations mentioned above is remarkable considering that our numerical setup differs considerably.

With the magnetic field amplified up to full saturation, albeit not yet in equipartition, we have used that configuration to observe its evolution into a quiescent disk by turning off supernova feedback, thereby removing the main driver of turbulence. We find that that with the decay of turbulence, the gas cools down and falls onto a thin clumpy disk with dense arms. Inside these arms, we observe the magnetic field to be locally at equipartition with turbulence and thermal pressure, with field strengths of several $\unit{\mu G}$. Field lines inside the quiescent disk are aligned with the disk plane, with a strong toroidal component and slightly weaker radial component.  These results are consistent with observations of spiral galaxies which usually have ordered magnetic field strengths of $5 \unit{\mu G}$ where the magnetic energy density is at equipartition with the turbulent energy density \citep{2016A&ARv..24....4B}. The toroidal as well as the radial components are symmetric across the midplane, with an average pitch angle of $\langle p\subscript{B} \rangle = -12\unit{^\circ}$. This symmetry confirms measurements of the Galactic magnetic field \cite{2012ApJ...755...21M} and pitch angle measurements of various nearby galaxies \cite{2015ApJ...799...35V}. Furthermore, the magnetic field reversals along the radial distance from the centre of the galaxy confirm observations of nearby spiral galaxies \citep{2016A&ARv..24....4B} where such reversals are found, usually attributed to their spiral structure.

\section{Conclusions}
\label{chap:conclusions}

We have performed numerical experiments with an idealised setup of a dwarf galaxy to study the evolution of a galactic magnetic seed field as small-scale dynamo amplification occurs due to turbulence driven by feedback processes, such as supernova explosions. We have shown that, with the formation and the death of the first massive stars, the gas swiftly becomes turbulent and the initially weak seed field grows exponentially. The e-folding time of this dynamo process becomes shorter and shorter as the numerical resolution increases. Since the Reynolds number in our simulations is several orders of magnitude lower than that of a typical ISM plasma, we extrapolate the small-scale dynamo efficiency to an e-folding time of $\tau = 1 \unit{Myr}$. We conclude that even an initially very weak field strength of $B = 10^{-20} \unit{G}$ would be amplified up to dynamo saturation in a time span of only $30 \unit{Myr}$. Thus, a newly born turbulent galaxy can be highly efficient in amplifying its early seed fields, whether they be primordial or generated during structure formation, extremely rapidly. This mechanism would establish considerably strong fields since the early stages of the Universe in turbulent galactic environments, which would explain the high magnetic field strengths found by \cite{M-2008Natur.454..302B} even at high redshift. The strong but sub-equipartition-strength field can then be further transformed by other processes like the $\alpha-\Omega$ dynamo. Moreover, the magnetic field will follow and even influence the history of its host galaxy through decisive events such as starbursts, mergers or more quiet phases. Therefore, magnetic fields should not be neglected when dealing with problems of galaxy evolution.

\section*{Acknowledgements}

This work was funded by the Swiss National Science Foundation SNF. 
All simulations were run on the Piz Dora cluster at the Swiss National Supercomputing Centre CSCS in Lugano, Switzerland.

\bibliography{michael,romain}

\end{document}